\documentclass[reqno,oneside,10pt]{amsart}

\usepackage{tikz}

\usepackage{amsmath,amsthm}
\usepackage{amsfonts,eucal}
\usepackage[psamsfonts]{amssymb}
\usepackage[applemac]{inputenc}
\usepackage[french]{babel}
\usepackage[T1]{fontenc}

\usepackage{mathptmx}
\usepackage{xspace}
\usepackage{setspace,geometry,multirow}
\usepackage[pdfpagemode=UseNone, pdfstartview={XYZ null null null}]{hyperref}
\hypersetup{
colorlinks=true,
citecolor=red,
linkcolor=blue,
urlcolor=blue
}
\usepackage[capitalise]{cleveref}

\newcommand{\picar}{\pi_{\, \begin{tikzpicture}[baseline={(current bounding box.center)},scale=0.13]\draw[line width=0.1mm] (0,0) -- (0,1) -- (1,1) -- (1,0) -- (0,0);\end{tikzpicture}}}
\newcommand{\pitri}{\pi_{\begin{tikzpicture}[baseline={(current bounding box.center)},scale=0.15]\draw[line width=0.1mm] (0,0) -- (1,0) -- (1/2,0.866) -- (0,0);\end{tikzpicture}}}
\newcommand{\pihex}{\pi_{\begin{tikzpicture}[baseline={(current bounding box.center)},scale=0.075]\draw[line width=0.1mm] (1., 0.) -- (0.5, 0.866025) -- (-0.5, 0.866025) -- (-1., 0.) --(-0.5, -0.866025) -- (0.5, -0.866025) -- (1,0);\end{tikzpicture}}}

\definecolor{rougePompier}{rgb}{0.93,0.11,0.14}
\definecolor{vertForet}{rgb}{0.04,0.55,0.07}

\newcommand{\invisible}[1]{} 

\title[Invariance conforme et universalit\'e]{L'invariance conforme et l'universalit\'{e}\\ au point critique des mod\`eles bidimensionnels}
\author[Yvan Saint-Aubin]{Yvan Saint-Aubin}
\address[Yvan Saint-Aubin]{
D\'{e}partement de math\'{e}matiques et de statistique\\
Universit\'{e} de Montr\'{e}al\\
Montr\'eal, QC, Canada, H3C 3J7.}
\email{yvan.saint-aubin@umontreal.ca}
\date{\today}

\begin{document}

\begin{abstract}
Des quelques articles publiés par Robert P.~Langlands en physique mathématique, c'est celui publié dans le {\it Bulletin of the American Mathematical Society} sous le titre {\it Conformal invariance in two-dimensional percolation} qui a eu, à ce jour, le plus d'impact : les idées d'Oded Schramm ayant mené à l'équation de Loewner stochastique et les preuves de l'invariance conforme de modèles de physique statistique par Stanislav Smirnov ont été suscitées, au moins en partie, par cet article. Ce chapitre rappelle sommairement quelques idées de l'article original ainsi que celles issues des travaux de Schramm et Smirnov. Il est aussi l'occasion pour moi de décrire la naissance de ma collaboration avec Robert Langlands et d'exprimer ma profonde gratitude pour cette fantastique expérience scientifique et humaine.

\bigskip

\noindent{\scshape Extended Abstract.} Of all mathematical physics contributions by Robert P.~Langlands, the paper {\it Conformal invariance in two-dimen\-sion\-al percolation} published in the {\it Bulletin of the American Mathematical Society} is the one that has had, up to now, the most significant impact: Oded Schramm's ideas leading to the stochastic Loewner equation and Stanislav Smirnov's proof of the conformal invariance of percolation and the Ising model in two dimensions were at least partially inspired by it. This chapter reviews briefly some ideas of the original paper and some of those by Schramm and Smirnov.

This chapter is also for me the occasion to reminisce about the extraordinary scientific and human experience that working with Robert Langlands was. It started in the late 80's when Langlands would spend Summers at the {\it Centre de recherches math\'ematiques} in Montreal. The ``Langlands program'' was already launched and many colleagues were devoting their career to it. Beside his steady efforts in automorphic forms, Langlands was already exploring new fields, mathematical physics being one of them. He studied conformal field theory, just then introduced, and started thinking about the renormalisation group. He presented some of these ideas in a study workshop in Montreal and this is when our collaboration took off. This collaboration concentrated on problems related to conjectures of universality and conformal invariance of two-dimensional discrete systems on compact domains, and on the Bethe Ansatz. Discussing, bouncing ideas and simply collaborating with Langlands was a fantastic experience. I had a hard time understanding his more formal presentations. But one-on-one discussions at the blackboard were always concrete, instructive and fruitful. My barrage of questions never seemed to frazzle him. Whenever he understood where I was blocked, his answer would often be ``Let me give you an example''. I had imagined that he would prefer the loftier way of mathematical communication through abstraction. But it was a nice surprise to discover that he knew so many concrete examples that revealed the crux of difficult mathematical concepts. I am deeply indebted to him for this collaboration that lasted about ten years and for his friendship that remains very much alive today.

\medskip

\end{abstract}

\maketitle

\tableofcontents

\onehalfspacing

\begin{section}{Prologue : réminiscences et gratitude}\label{sec:intro}

Robert P.~Langlands commença à visiter le Centre de recherches mathématiques à Montréal durant les étés des années quatre-vingt. Le programme de Langlands, à l'intersection des théories des formes automorphes, de la représentation et de l'analyse, était déjà enclenché et plusieurs mathématiciens y travaillaient. Le présent recueil y est principalement consacré. Sans l'intention de délaisser ses premiers amours, l'aventurier Langlands était curieux d'affronter de nouveaux défis. Il explora d'autres domaines, loin des formes automorphes, dont certains problèmes liés à la physique. Il consacra d'ailleurs à ses nouvelles explorations la Chaire Aisenstadt du Centre de recherches mathématiques qu'il détint en 1988-1989. Notre intérêt commun pour les théories des champs conformes qui venaient d'être proposées aurait pu être le point de départ d'une collaboration. Mais ce sont ses efforts pour exposer, lors d'ateliers d'étude estivaux, les idées de Kesten sur la percolation et les siennes sur le groupe de renormalisation qui lancèrent nos discussions. Pendant les dix années qui suivirent, nous travaillâmes sur des problèmes liés aux hypothèses d'universalité et d'invariance conforme, principalement sur des domaines compacts, ainsi que sur l'Ansatz de Bethe. Je ne suis pas le seul à avoir profité de cette collaboration : des étudiants de l'Université de Montréal, surtout parmi les miens, ont eu la chance de travailler avec lui.

Discuter, échanger des idées et collaborer avec Robert Langlands fut une expérience fantastique. Plusieurs collègues prirent le soin, au début, de m'informer de la prestance scientifique de mon nouveau collaborateur. Je devrais leur être reconnaissant, mais ma relative ignorance des formes automorphes n'obscurcissait pas complètement mon panorama mental des mathématiques de la seconde moitié du vingtième siècle. Les exposés de Robert ne font que faire miroiter ses buts et les chemins qu'il pressent pour les atteindre. Après plus d'un, j'ai été pris du doute de n'avoir contemplé qu'un mirage. Mais nos discussions, seul à seul, furent toujours concrètes, instructives et fertiles. Jamais le déluge de mes questions n'a été reçu par une quelconque impatience. Et, lorsque Robert comprenait l'endroit où je butais, sa réponse était souvent, à ma grande surprise : \og Je te donne un exemple ! \fg. Chaque membre de notre communauté possède sa personnalité mathématique. J'avais pensé que celle de Robert se limitait à penser et à communiquer les mathématiques en les termes les plus abstraits. Ce fut donc une révélation que de reconnaître que ses outils ne se limitent pas à l'abstraction, mais qu'ils incluent moults exemples bien tangibles. Je lui suis donc profondément reconnaissant pour ces échanges, parfois abstraits parfois concrets, pour cette collaboration fructueuse de près de dix ans, et pour cette chaleureuse amitié qui, elle, demeure toujours vivante.

J'ai choisi de ne discuter qu'un seul des articles que Langlands a écrits en physique mathématique. En fait, cette étude n'occupera même qu'une partie des pages qui me sont allouées. La raison de ce choix est simple : cet article, {\itshape Conformal invariance in two-dimensional percolation} \cite{LPS}, est son article dans nos domaines d'intérêts communs qui a eu le plus grand impact en mathématiques et en physique. Modestement, je crois que ce texte a permis d'éclairer sous un angle nouveau des sujets déjà bien établis dans les deux communautés, la percolation et les transitions de phase, suffisamment pour que de jeunes mathématiciens audacieux y percent de nouvelles avenues. La première section sera donc consacrée à cet article que nous avons écrit avec Philippe Pouliot, les deux suivantes aux travaux pionniers d'Oded Schramm et de Stanislav Smirnov, respectivement.

\end{section}
\begin{section}{Deux hypothèses sur les probabilités de traversée critiques}\label{sec:un}

Soit $D\subset \mathbb R^2$ un domaine connexe de frontière $\partial D$ le long de laquelle deux segments disjoints $I$ et $J$ sont choisis. Dans ce même plan $\mathbb R^2$, les points $\delta (m,n)$ où $m,n\in\mathbb Z$ et $\delta\in(0,\infty)$ sont identifiés au graphe $\mathbb Z^2$ dont les arêtes lient les voisins immédiats. La maille $\delta$ sera choisie suffisamment petite pour qu'aucune arête du plongement de $\mathbb Z^2$ dans $\mathbb R^2$ n'intersecte simultanément les deux segments $I$ et $J$ et l'intersection $D\cap \mathbb Z^2$ soit un graphe connexe. Seuls les points de $\mathbb Z^2$ à l'intérieur de $D$ et ceux liés par une arête à un sommet à l'intérieur joueront un rôle. Une {\em configuration} de ces points {\em intérieurs} est obtenue en donnant à chacun de ces points un statut, soit {\em ouvert}, soit {\em fermé}. Si $p\in[0,1]$ est la probabilité qu'un de ces sommets intérieurs soit ouvert, alors la configuration se voit accorder la probabilité $p^{\textrm{\#(sommets ouverts)}} (1-p)^{\textrm{\#(sommets fermés)}}$. Une {\em traversée entre $I$ et $J$ à l'intérieur de $D$} existe si les conditions suivantes sont remplies : d'abord il existe une arête chevauchant le segment $I$ et une le segment $J$ dont les extrémités soient des sommets ouverts, puis il est possible de joindre ces arêtes chevauchant $I$ et $J$ par des arêtes contiguës dont les extrémités sont ouvertes. L'existence d'une telle traversée dépend évidemment de la configuration. Il est donc naturel de définir la {\em probabilité $\pi(D,I,J;\delta;p)$ d'une traversée} allant de $I$ à $J$ à l'intérieur de $D$; c'est donc la somme des probabilités des configurations possédant une telle traversée. Cette définition rappellera le passage d'un liquide dans un milieu aléatoire qui donne le nom \og percolation \fg{} à ce domaine des probabilités. Dans cette interprétation, le paramètre $\delta$ mesure les dimensions relatives du réseau $\mathbb Z^2$ (de la mouture du café) et du domaine $D$ (le filtre où l'eau percole). Cette fonction $\pi(D,I,J;\delta;p)$ est un polynôme en $p$ définissant une bijection de l'intervalle $[0,1]$ sur lui-même. Le diagramme gauche de la figure \ref{fig:un} présente un domaine $D$ carré avec les deux segments $I$ et $J$ de sa frontière identifiés par un trait gras. La configuration choisie possède une traversée de $I$ à $J$, indiquée en pointillés. Cependant, cette même configuration n'a pas de traversée entre les segments disjoints $I'$ et $J'$ dont l'union est le complément dans la frontière de $I\cup J$.   
\begin{center}
\begin{figure}
\includegraphics[width = .45\linewidth]{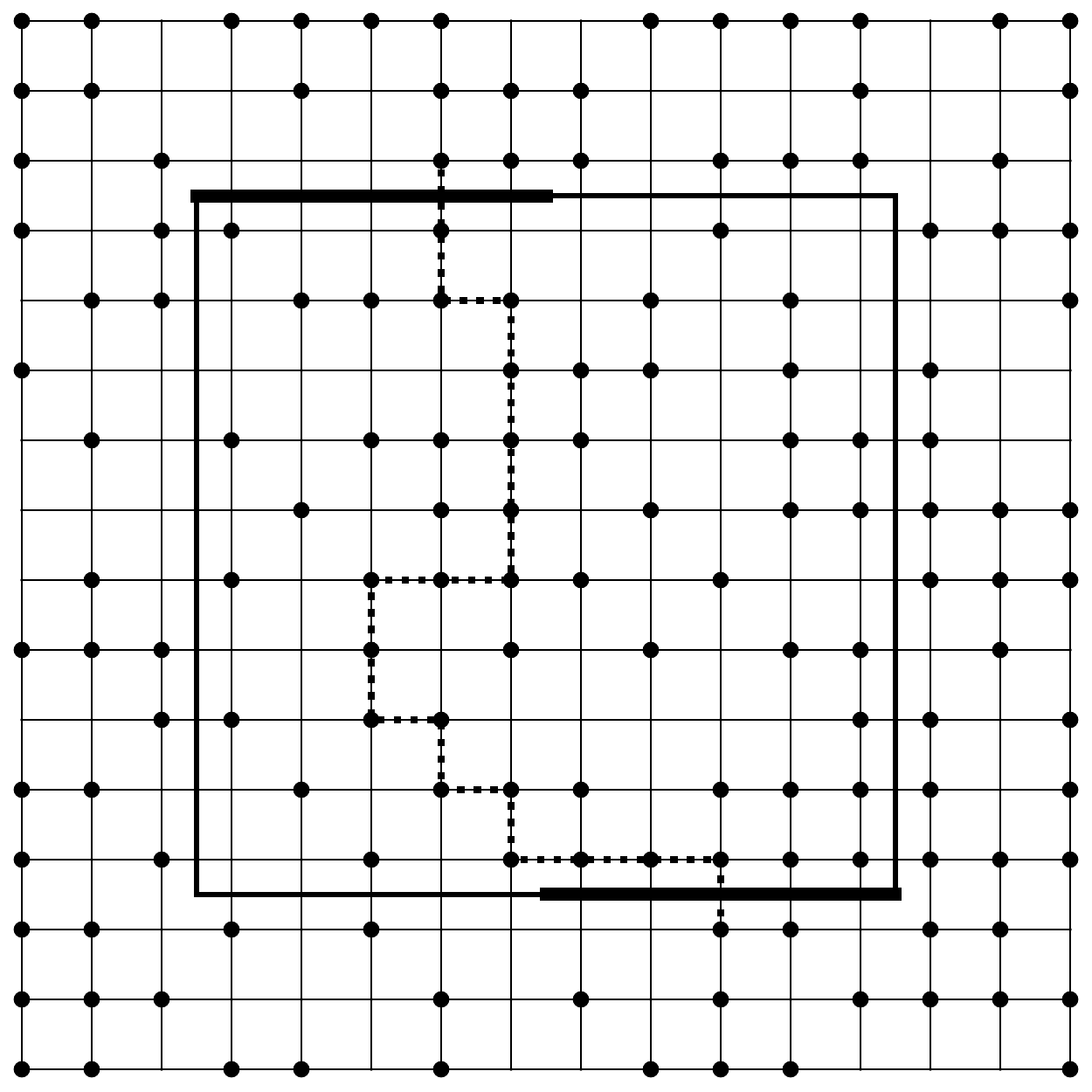}\hfill
\includegraphics[width = .45\linewidth]{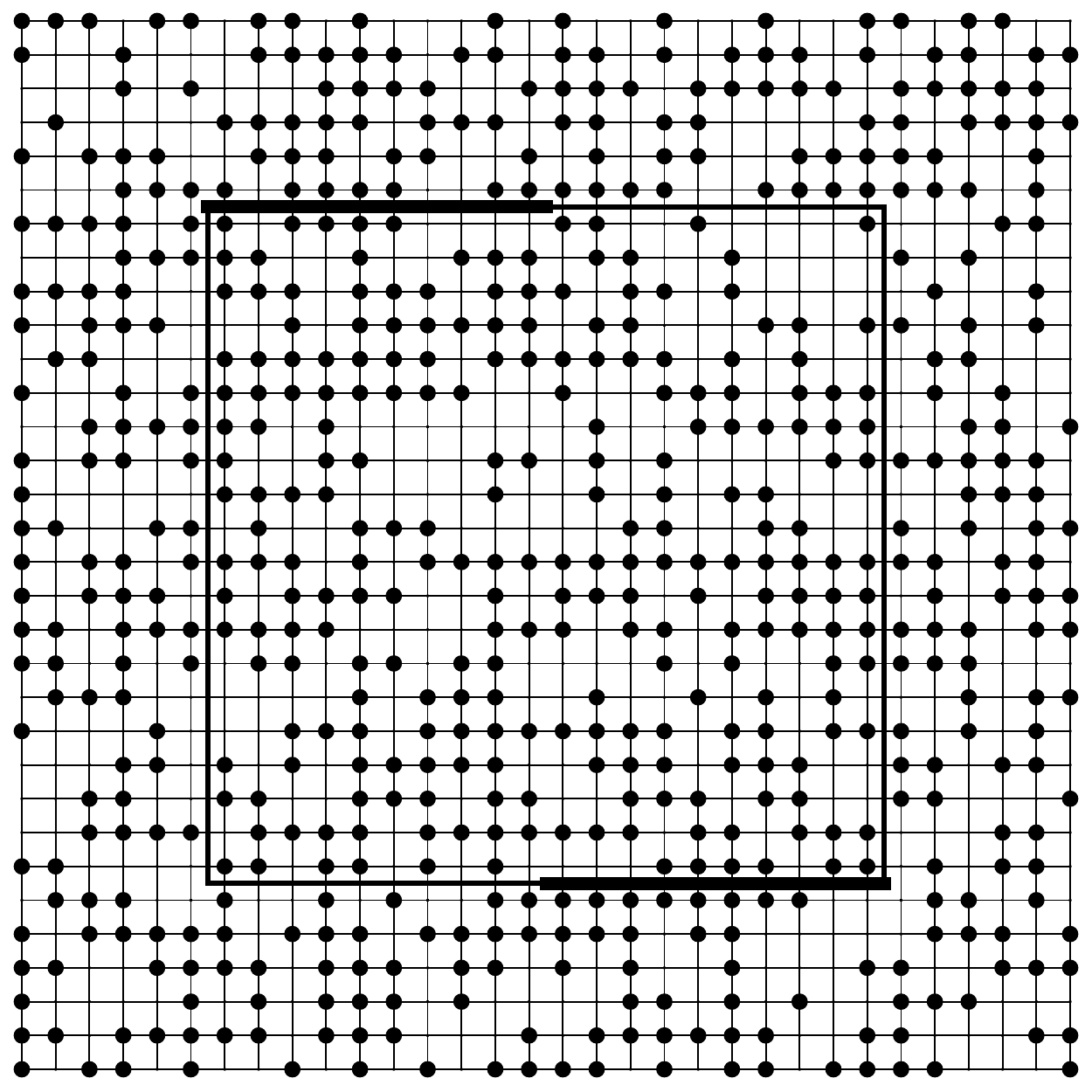}
\caption{Deux configurations pour le même domaine $D$, mais deux mailles distinctes. Les deux configurations contiennent une traversée de $I$ à $J$; celle du diagramme de gauche est indiquée en pointillés.}\label{fig:un}
\end{figure}
\end{center}

L'étude de ces probabilités $\pi(D,I,J;\delta;p)$ est malaisée et d'intérêt limité. Notre objet d'étude sera plutôt la limite $\pi(D,I,J;p)=\lim_{\delta\to 0}\pi(D,I,J;\delta;p)$ quand la maille du réseau $\mathbb Z^2$ tend vers zéro, alors que le domaine $D$ est tenu fixe. Le diagramme droit de la figure \ref{fig:un} décrit le même domaine, mais sur lequel est superposé un plongement de $\mathbb Z^2$ avec une maille plus fine que celle du diagramme de gauche. Il est clair que d'autres définitions, pour les conditions de départ et d'arrivée de la traversée, pour la position relative du domaine par rapport au réseau plongé, pour le processus limite lui-même, etc., pourraient être choisies. Ces précisions s'avéreront avoir peu d'importance pour notre propos. Kesten \cite{Kesten} (voir également \cite{AB}) a montré qu'il existe une probabilité $p_c\in(0,1)$, dite {\em critique}, telle que
\begin{equation}\label{eq:kesten1}\pi(D,I,J;p)=\begin{cases}0, & p<p_c,\\
                            1, & p>p_c,\end{cases}
\end{equation}
et telle que 
\begin{equation}\label{eq:kesten2}0<\liminf_{\delta\to 0}\pi(D,I,J;\delta;p_c)\leq \limsup_{\delta\to 0}\pi(D,I,J;\delta;p_c)<1.\end{equation}
Le résultat de Kesten laisse ouverte la détermination de la limite en $p_c$. En fait il n'est pas clair qu'elle existe ou que son existence soit indépendante des diverses variations des définitions ci-dessus. Mais, intuitivement, si cette limite existe, elle devrait dépendre du domaine $D$ et des segments $I$ et $J$ choisis. Ainsi les probabilités de traversée horizontale entre les côtés opposés d'un rectangle devraient dépendre du rapport des longueurs de ses côtés. Par exemple, il semble raisonnable de supposer que, en $p_c$, la probabilité d'une traversée horizontale dans le rectangle à gauche de la figure \ref{fig:deux} devrait être plus grande que celle dans le rectangle de droite.
\begin{center}
\begin{figure}
\includegraphics[width = .45\linewidth]{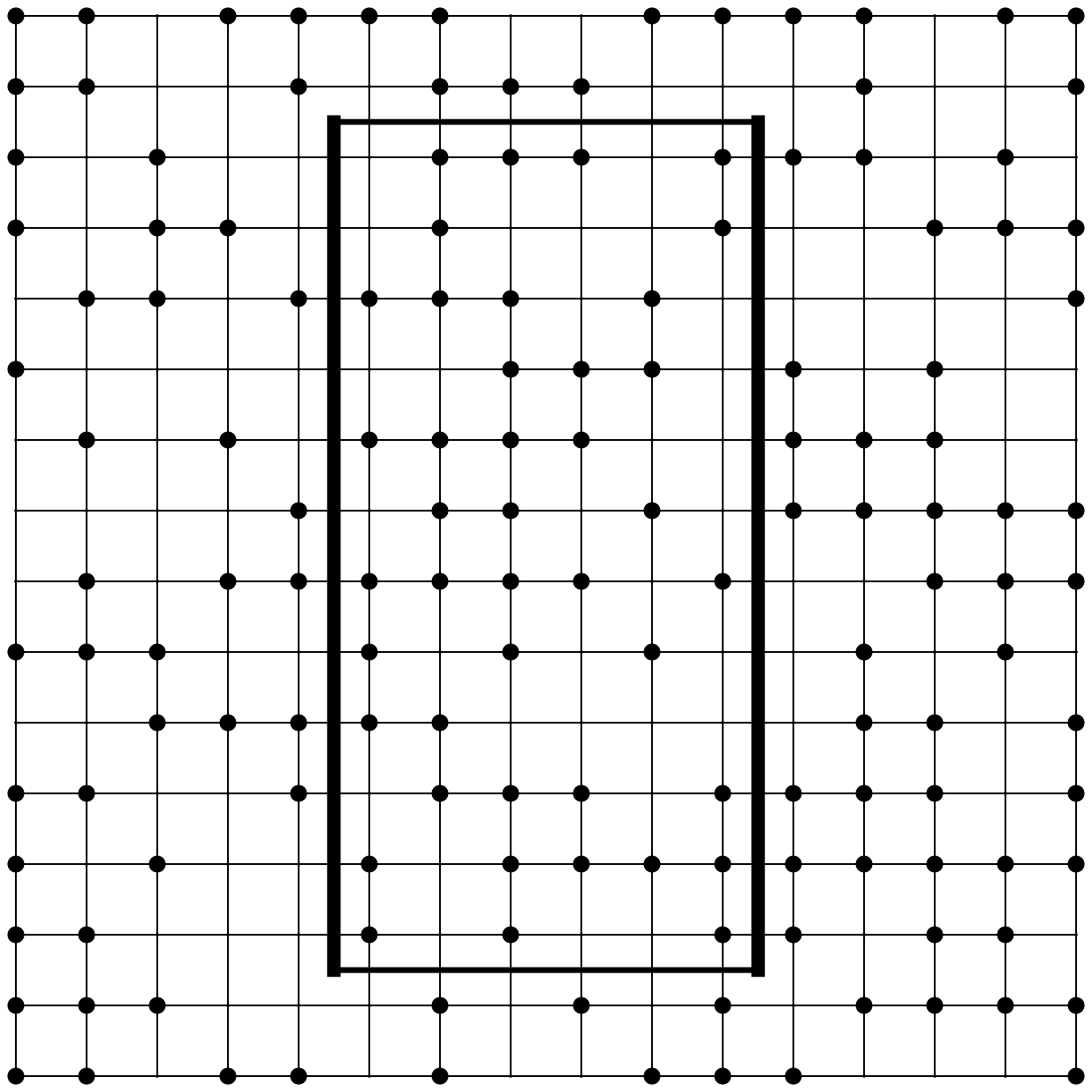}\hfill
\includegraphics[width = .45\linewidth]{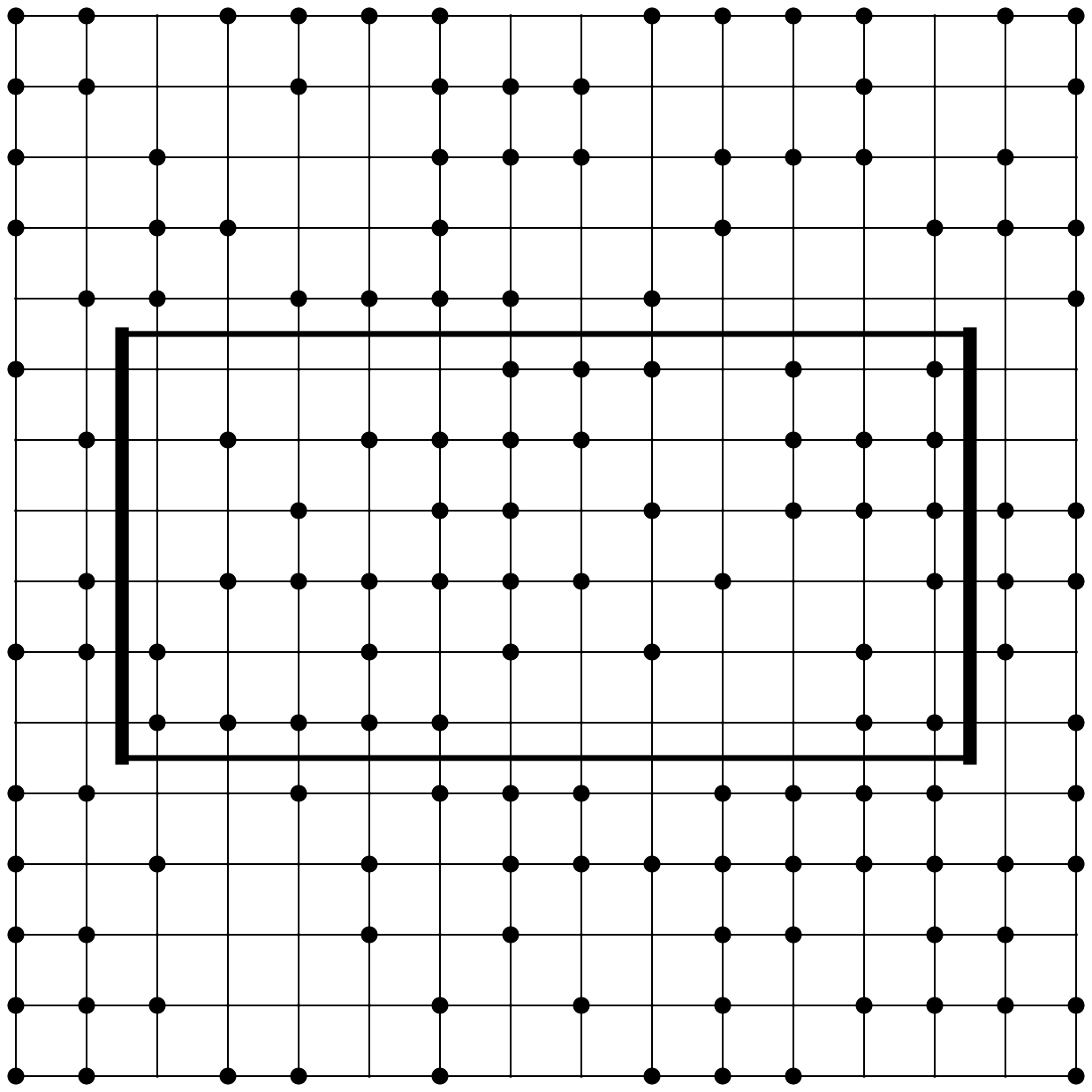}
\caption{La probabilité de traversée horizontale dans le domaine rectangulaire de gauche devrait être supérieure à celle du domaine de droite.}\label{fig:deux}
\end{figure}
\end{center}

Concentrons-nous pour l'instant sur les domaines rectangulaires, comme ceux de la figure \ref{fig:deux} où les côtés gauche et droit sont les segments $I$ et $J$. Notons par $r$ le rapport de la longueur des côtés verticaux sur celle des horizontaux. Supposons enfin que les limites $\lim_{\delta\to 0}\pi(D,I,J;\delta;p_c)$ existent. Les probabilités de traversée ont été introduites ci-dessus sur le réseau carré $\mathbb Z^2$. Mais tout autre réseau périodique aurait pu être utilisé. En fait, notre formulation \eqref{eq:kesten1} et \eqref{eq:kesten2} du résultat de Kesten ne lui rend pas justice : il a démontré ce résultat pour une large famille de graphes périodiques définis comme suit. Un graphe $\mathcal G$, plongé dans $\mathbb R^2$, est dit {\em périodique} si 
\begin{itemize}
\item[{\em (i)}] $\mathcal G$ ne contient pas de boucles (au sens des graphes);
\item[{\em (ii)}] $\mathcal G$ est périodique sous translation par les éléments d'un réseau $L\subset \mathbb R^2$ de rang deux;
\item[{\em (iii)}] le nombre d'arêtes attachées à un site est borné;
\item[{\em (iv)}] la longueur d'une arête est finie et tout ensemble compact de $\mathbb R^2$ intersecte un nombre fini d'arêtes et
\item[{\em (v)}] $\mathcal G$ est connexe.
\end{itemize}
Clairement les réseaux carré, triangulaire et hexagonal sont périodiques dans ce sens.
Même si ses résultats valent hors $p_c$, ils soulèvent la question naturelle : les probabilités 
$\picar(r)$,
$\pitri(r)$ et 
$\pihex(r)$
 de traversée horizontale en $p_c$, à l'intérieur d'un rectangle de rapport $r$, sur les réseaux carré, triangulaire et hexagonal, sont-elles reliées entre elles ? C'est d'un débat sur cette question qu'est née notre première collaboration : Langlands arguait que ces trois fonctions $\picar(r)$, $\pitri(r)$ et $\pihex(r)$ devaient coïncider, alors que je pensais que ces probabilités pouvaient appartenir à des classes d'universalité distinctes. (Ces positions reposaient sur notre compréhension des arguments physiques liés à la description des phénomènes critiques par la théorie des champs conformes. Dans cette théorie, les phénomènes critiques, telles les traversées horizontales en $p_c$, sont groupés en classes d'universalité étiquetées, entre autres, par la valeur de l'élément central $c$ de l'algèbre de Lie qui gouverne ces théories, l'algèbre de Virasoro. Ces arguments physiques associaient à la percolation la valeur $c=0$.) Notre premier article \cite{LPPS}, écrit conjointement avec C.~Pichet et P.~Pouliot, présente des simulations sur ordinateur déterminant numériquement ces probabilités de traversée sur des rectangles pour plusieurs rapports $r$ sur les trois réseaux ci-dessus ainsi que pour trois autres modèles de percolation où ce sont les arêtes qui sont ouvertes ou fermées, plutôt que les sommets. Ces simulations nous convainquirent que ces limites existent et que, tel que Langlands l'avait pressenti, ces probabilités vues comme fonction de $r$ sont égales pour les six modèles considérés. Puisque toutes ces simulations étaient menées sur un réseau fini (et donc pour une maille plus grande que zéro), nous dûmes proposer une approximation des probabilités critiques $p_c$ propices aux mesures à faire. (Même si les probabilités de traversée $\picar$, $\pitri$, ..., coïncident, les probabilités critiques $p_c$ dépendent, elles, du réseau. La notice {\em percolation threshold} sur l'encyclopédie en ligne Wikipedia collige les meilleures valeurs connues de ces probabilités critiques pour une multitude de réseaux, périodiques ou non, en deux dimensions ou plus.) Plusieurs des techniques que nous utilisâmes étaient probablement connues, mais la coïncidence des fonctions probabilités de traversée était l'observation nouvelle. Même si ce premier travail observe numériquement l'universalité de ces probabilités de traversée, la formulation générale d'une hypothèse d'universalité n'apparaît dans toute sa généralité que dans notre second travail \cite{LPS}, écrit conjointement avec P.~Pouliot.
 
S'il est possible de changer le réseau sur lequel la percolation a lieu, il est également possible de changer la probabilité que telle ou telle partie du système soit ouverte. Par exemple, pour le réseau $\mathbb Z^2$, les sites dont la somme des coordonnées  est paire pourraient être ouverts avec une probabilité $p_p$ distincte de celle, disons $p_i$, des sites où cette somme est impaire. Soit donc une fonction $p:\mathcal S\to [0,1]$ de l'ensemble des sites $\mathcal S$ du graphe périodique $\mathcal G$ qui soit périodique sous les translations du réseau $L$ de la définition {\em (ii)} ci-dessus. Comme précédemment, les probabilités de traversée peuvent être définies sur ce graphe où chaque site peut avoir une probabilité d'être ouvert distincte de ces voisins. Cette paire $M=M(\mathcal G, p)$ définit donc un nouveau modèle de percolation et nous noterons par $\pi_M$ les probabilités de traversée qui s'y rattachent. Puisque $\mathcal G$ est plongé dans $\mathbb R^2$, tout élément $g$ de $GL(2,\mathbb R)$ transforme ce graphe périodique en un graphe périodique $\mathcal G'=g\mathcal G$ par la simple action $s\to gs$ sur les sites et similairement sur les arêtes. Soit donc $gM$ le modèle défini sur ce nouveau graphe. Si la même transformation linéaire est appliquée au domaine $D$ et aux segments $I$ et $J$, les probabilités de traversée vérifient trivialement la relation $\pi_M(D,I,J)=\pi_{gM}(gD,gI,gJ)$. Cependant, les probabilités $\pi_M(D,I,J)$ et $\pi_{gM}(D,I,J)$ sont en général différentes. L'hypothèse d'universalité relie les modèles $M$ et $M'$ associés à deux paires $(\mathcal G,p)$ et $(\mathcal G',p')$ distinctes.

\smallskip

\noindent{\bfseries Hypothèse d'universalité} --- {\em Soient $M$ et $M'$ deux modèles de percolation critiques. Il existe un élément $g\in GL(2,\mathbb R)$ tel que
$$\pi_{gM}(D,I,J)=\pi_{M'}(D,I,J)$$
pour tous les domaines $D$ et segments $I$ et $J$.}

\smallskip

\noindent La vérification numérique offerte par notre premier article \cite{LPPS} est directe, mais quelque peu limitée : les réseaux utilisés sont plongés de façon régulière, c'est-à-dire de façon à préserver l'invariance sous rotation autour des sommets par un angle de $\frac{\pi}2$, $\frac{\pi}3$ et $\frac{2\pi}3$ pour les réseaux carré, triangulaire et hexagonal respectivement, et l'élément $g\in GL(2,\mathbb R)$ faisant le lien entre les modèles étudiés est toujours (un multiple de) l'identité. Le second article donne cependant un exemple d'un modèle $M(\mathcal G,p)$ où le graphe est $\mathbb Z^2$, mais $p$ n'est pas la fonction constante. Pour ce modèle $M(\mathcal G,p)$, l'élément $g$ le reliant au modèle sur le réseau carré usuel n'est ni diagonal, ni une matrice orthogonale. 

Il est probable que cet énoncé vaille pour des modèles qui ne sont pas définis à l'aide d'un réseau périodique dans le sens de Kesten : viennent à l'esprit des modèles définis à partir des pavages non périodiques de Penrose ou même ceux dont le graphe est lui-même aléatoire. Quoiqu'il en soit, la conjecture demeure à ma connaissance ouverte.

Deux collègues, M.~Aizenman et J.~Cardy, nous ont aidés à façonner la seconde hypothèse et à en faire un défi mathématique excitant. Après que nos premières données numériques sur les probabilités de traversée eurent été rendues publiques, Langlands eut la chance de discuter avec Aizenman qui lui proposa comment l'invariance conforme pouvait peut-être se manifester. À cette époque, la théorie des champs conformes avait déjà montré sa puissance pour prédire les exposants critiques qui caractérisent les transitions de phase en deux dimensions en leurs points critiques, et pour obtenir des formes explicites des fonctions de corrélation de diverses quantités physiques de ces modèles statistiques. Malgré que la percolation était parmi les modèles physiques étudiés par cette théorie des champs conformes, les probabilités de traversée étaient fort loin des quantités qu'elle avait permis de décrire. C'est pourquoi j'avais été 
étonné par l'audacieuse suggestion d'Aizenman. 

Soit $j$ une transformation linéaire du plan $\mathbb R^2$ dont le carré est moins l'identité. Elle définit donc une structure complexe sur le plan et la notion de fonctions $j$-holomorphes. Si le triplet $(D,I,J)$ est la donnée d'un domaine et de deux segments disjoints le long de sa frontière, une fonction $\phi:\mathbb R^2\to\mathbb R^2$ qui est $j$-holomorphe à l'intérieur de $D$, et continue et bijective jusqu'à sa frontière, définit un nouveau triplet $(\phi(D),\phi(I),\phi(J))$ d'un domaine avec deux segments disjoints $\phi(I)$ et $\phi(J)$ le long de sa frontière. Cette observation vaut également si $\phi$ est $j$-antiholomorphe à l'intérieur de $D$.

\smallskip

\noindent{\bfseries Hypothèse d'invariance conforme} --- {\em Pour tout modèle de percolation critique $M=M(\mathcal G,p)$, il existe une transformation linéaire $j$ définissant une structure complexe telle que
$$\pi_M(\phi(D),\phi(I),\phi(J))=\pi_M(D,I,J)$$
pour tout triplet $(D,I,J)$ et pour tout $\phi$ qui soit $j$-(anti)holomorphe à l'intérieur de $D$, continue et bijective jusqu'à sa frontière.
}

\smallskip
\begin{center}
\begin{figure}[h!]
\includegraphics[width = .45\linewidth]{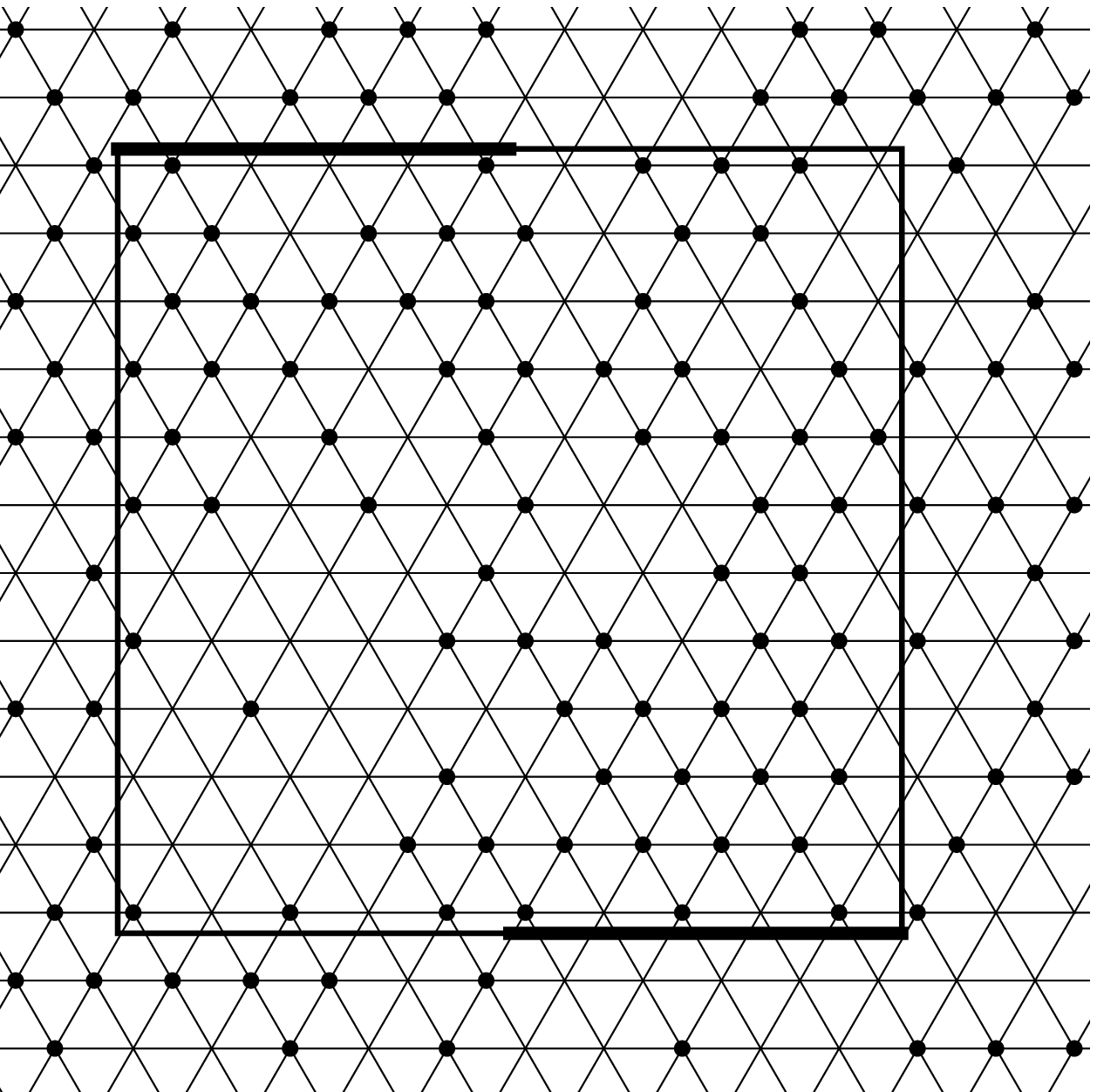}\hfill
\includegraphics[width = .45\linewidth]{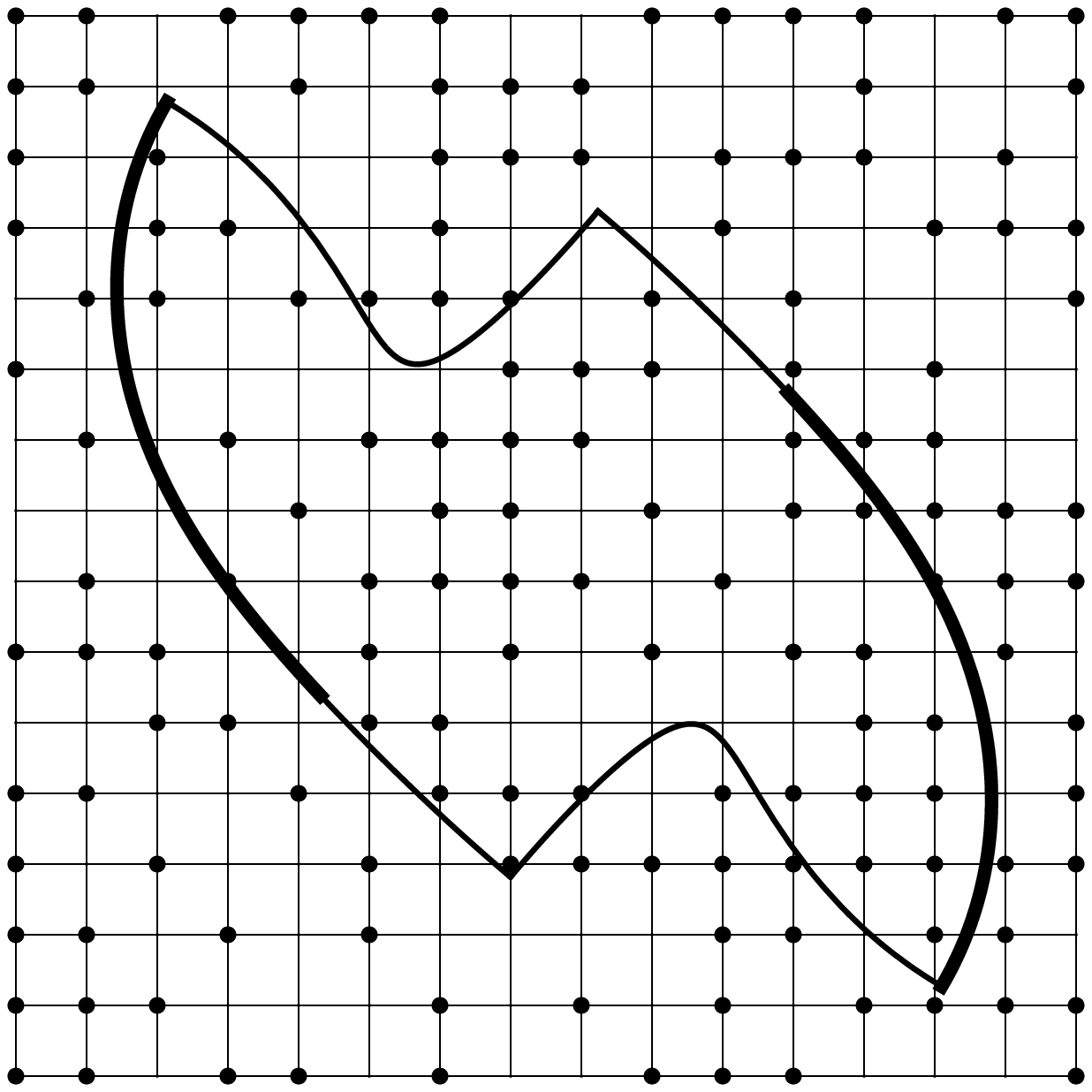}
\caption{Deux probabilités de traversée qui devraient être égales, dans la limite $\delta\to0$, à celle du triplet de la figure \ref{fig:un} : celle de gauche sur le réseau triangulaire (hypothèse d'universalité) et celle de droite pour un triplet $(D',I',J')$ obtenu de celui utilisé à la figure \ref{fig:un} par une fonction holomorphe (hypothèse d'invariance conforme).
}\label{fig:trois}
\end{figure}
\end{center}

\noindent Cette hypothèse veut dire en particulier que la probabilité de traversée pour le triplet $(D',I',J')$ présenté au diagramme droit de la figure \ref{fig:trois} devrait être égale, dans la limite $\delta\to 0$, à celle du carré utilisé à la figure \ref{fig:un}. Malgré son audace, l'hypothèse est possiblement vraie pour une famille plus grande de fonctions. Notre second article \cite{LPS} présente une vérification numérique pour la fonction complexe $z\to z^2$ appliquée à un domaine $D$ carré incluant l'origine. Sur un tel domaine, cette fonction n'est certainement pas bijective. Pour contrer cette difficulté, le réseau utilisé dans la mesure numérique est une paire de graphes $\mathbb Z^2$ joints le long d'une coupure longeant les points dont la première coordonnée est positive et la seconde nulle. On peut voir intuitivement cette paire de graphes $\mathbb Z^2$ comme couvrant le double recouvrement $\mathbb X$ du plan complexe qui rend bijective la fonction $\phi:\mathbb C\to\mathbb X$ donnée par $z\to z^2$. Sur ce \og double recouvrement \fg, l'hypothèse d'invariance conforme est remarquablement vérifiée.

Fort de ces observations numériques, Langlands contacta Cardy, lui demandant s'il était possible d'extraire, de la théorie des champs conformes, une prédiction théorique pour ces probabilités de traversée. La réponse fut rapide et convaincante. Maintenant connue sous le nom de {\em formule de Cardy}, la probabilité de traversée prédite par cette théorie des champs conformes est
\begin{equation}\label{eq:cardy}
\picar(r)=\frac{3\Gamma(\frac23)}{\Gamma(\frac13)^2}\sin^{\frac23}(\theta(r))\,_2F_1\left({\textstyle\frac13},{\textstyle\frac23};{\textstyle\frac43};\sin^2(\theta(r))\right)
\end{equation}
où $\Gamma$ et $\,_2F_1$ sont les fonctions gamma et hypergéométrique usuelles, et $\theta(r)$ est déterminé comme suit. Par le théorème de Riemann, il existe une fonction holomorphe envoyant l'intérieur du rectangle de rapport de côtés $r$ vers l'intérieur du disque de rayon $1$. Cette fonction peut être choisie de façon à ce que les sommets du rectangle aient pour images les point $z=e^{i\theta}$, $\bar z$, $-z$ et $-\bar z$. Cet angle $\theta=\theta(r)$ est celui apparaissant dans la formule ci-dessus. La courbe de la figure \ref{fig:cardy} montre cette prédiction de Cardy à laquelle nous avons ajouté les mesures pour les probabilités de traversée sur le réseau carré obtenues pour des rectangles contenant approximativement un million de sites. L'échantillon choisi est tel que l'erreur sur ces mesures est plus petite que la grosseur des points sur la figure. On s'inquiètera peut-être du léger écart entre la prédiction et les mesures pour les valeurs de $r$ extrêmes; cet écart est probablement dû au fait que, pour ces valeurs de $r$, le nombre de sites d'un des côtés du rectangle était trop petit pour atteindre la qualité de mesure des autres points. L'accord saisissant entre la prédiction et les mesures a été probablement un des éléments attirant de jeunes mathématiciens à ces conjectures.
\begin{center}
\begin{figure}[h!]
\includegraphics[width = .6\linewidth]{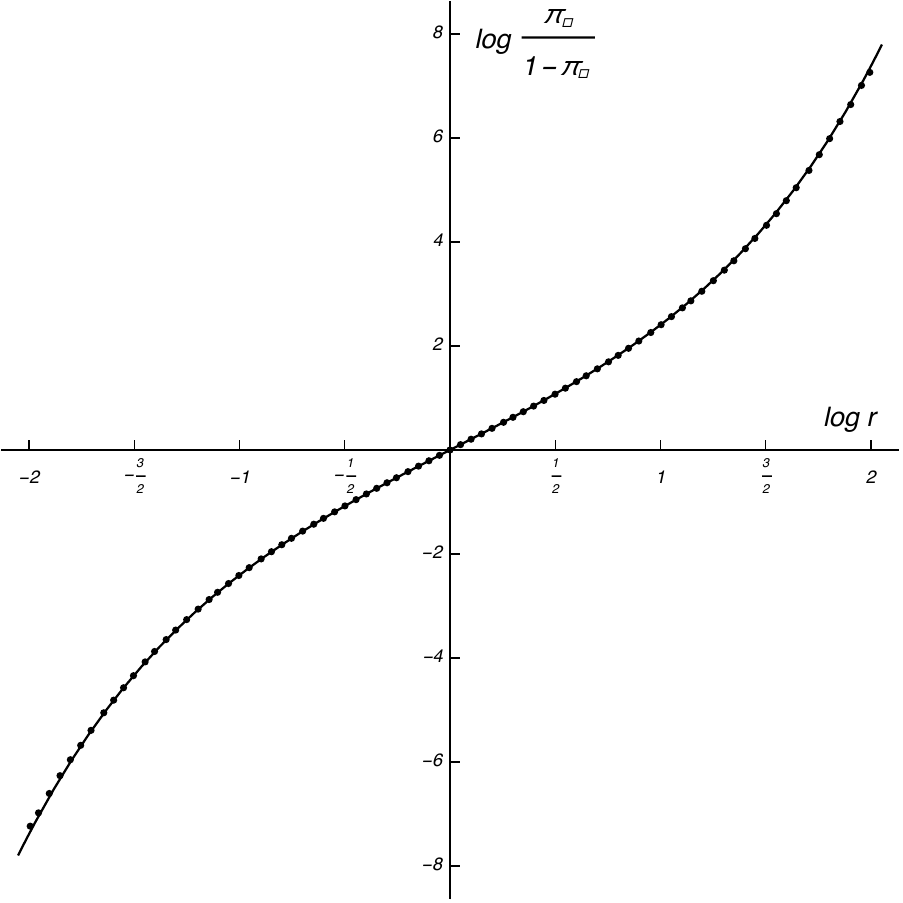}\hfill
\caption{La prédiction de Cardy (trait continu) avec les mesures numériques pour le réseau carré de \cite{LPS}.
}\label{fig:cardy}
\end{figure}
\end{center}

\noindent Il existe une forme remarquablement simple de la formule de Cardy que Smirnov \cite{Smirnov} attribue à L.~Carleson. Elle est donnée pour le triplet $(D,I,J)$ où $D$ est un triangle équilatéral de côté unité, $I$ un des côtés et $J$ est le segment le long d'un des deux autres côtés commençant au sommet opposé à $I$ et de longueur $x$. Alors $\pi(D,I,J)=x$.

Avant de clore cette section, voici quelques mots sur une autre avenue explorée par Langlands. Une des idées-clés de la physique statistique et de la théorie des champs quantiques est de remplacer la description ab initio d'un phénomène donné en \og intégrant \fg{} les interactions à petite échelle, ne gardant ainsi que le comportement à plus grande échelle. Cette opération, nommée {\em renormalisation}, est souvent décrite par son action sur les divers paramètres décrivant la théorie. Par exemple, les atomes d'un cristal sont organisés sur un réseau de maille $\delta$. Un ensemble de paramètres est alors nécessaire pour décrire l'interaction entre deux atomes à distance $\delta$, entre deux atomes à distance $2\delta$, et ainsi de suite. Les points fixes sous la renormalisation sont liés aux points critiques du phénomène. L'étude rigoureuse de cette opération de renormalisation, si elle est possible, requiert usuellement l'étude d'un nombre infini de paramètres. Une des idées de Langlands était de mener (rigoureusement) cette étude de la renormalisation sur une famille de modèles, chacun possédant un nombre fini $N$ d'états ou d'\og événements \fg{}. Dans ses modèles l'opération de renormalisation agit sur le cube $[0,1]^N$ dont les coordonnées sont les probabilités de chacun des événements.

Pour comprendre les modèles qu'il introduisit \cite{dualitat, marcAndre}, les triplets $(D,I,J)$ sont remplacés par les événements $E=(D,(I_{i_1},I_{j_1}),(I_{i_2},I_{j_2}),\dots, (I_{i_n},I_{j_n}))$. Tous ces {\em événements} $E$ auront lieu à l'intérieur d'un domaine $D$ carré. Cependant, les côtés sont maintenant divisés en $m$ segments, produisant ainsi $4m$ segments disjoints $I_1,I_2,\dots, I_{4m}$. Parmi ces segments disjoints, certaines paires $(I_{i_1},I_{j_1}),(I_{i_2},I_{j_2}),\dots, (I_{i_n},I_{j_n})$ sont choisies et $E$ est caractérisé par l'exigence, pour chacune de ces paires, qu'elle soit ou ne soit pas reliée. Par exemple, un événement $E$ pourrait ne contraindre que deux paires :  la paire $(I_3, I_5)$ devrait être reliée par des sites ouverts mais la paire $(I_3, I_8)$ devrait ne pas l'être; les autres paires sont libres d'être ou non reliées. L'opération de renormalisation introduite par Langlands regroupe alors $k^2$ de ces événements pour former un nouveau carré avec $4mk$ segments à sa frontière. Des règles sont alors données pour fusionner ces segments par ensembles de $k$ segments contigus, redonnant ainsi un événement décrit par $4m$ segments. Les résultats numériques présentés dans \cite{marcAndre} indiquent que ces modèles finis possèdent un point critique et qu'un des exposants critiques en ce point est proche de celui prédit pour la percolation. Clairement ces modèles finis devraient être explorés plus avant.

\end{section}

\begin{section}{L'équation de Schramm-Loewner}

Les hypothèses d'invariance conforme et d'universalité telles que formulées dans notre article dans le {\em Bulletin of the AMS} n'étaient que des conjectures dont les conséquences demeuraient à être explorées. Deux pionniers, Oded Schramm et Stanislav Smirnov, se mirent rapidement à pied d'\oe{}uvre pour explorer, formuler et prouver mathématiquement ces hypothèses. Les travaux de Schramm répondent à la question : existe-t-il des espaces de probabilité, permettant de mesurer les probabilités de traversée et respectant une condition d'invariance conforme similaire à celle que nous avions formulée ? Sa réflexion le mena à lier l'existence de tels espaces à celle du mouvement brownien. Le but de la présente section est modeste : il est d'expliquer, au moins intuitivement, ce lien.
La présentation ci-dessous tient donc pour acquis qu'un espace de probabilité existe capturant les propriétés des plages de sites ouverts et de sites fermés et permettant de répondre à la question : quelle est la probabilité d'une traversée pour le triplet $(D,I,J)$ en $p=p_c$ ? 
De plus, notre réflexion pourra utiliser le modèle de percolation $M=M(\mathcal G,p)$ de notre choix puisque l'hypothèse d'universalité suppose que l'espace de probabilité obtenu à la limite n'en dépend pas.

Les probabilités de traversée ont été introduites sur des domaines compacts, tels les rectangles, mais si l'hypothèse d'invariance conforme tient, ces probabilités de traversée devraient être reliées aux probabilités d'aller d'un segment à un autre de la frontière du demi-plan supérieur. Plutôt que de tracer une configuration de sites ouverts ou fermés, nous nous concentrerons maintenant sur une seule interface, c'est-à-dire la courbe séparant une plage de sites ouverts de la plage de sites fermés contiguë. 

La figure \ref{fig:cinq} présente deux interfaces dans un domaine carré (deux diagrammes du haut) et leur image sous l'application conforme envoyant ce domaine sur le demi-plan supérieur (les deux du bas). Ces interfaces démarrent au sommet $A$ du domaine carré. Puisque $p_c>0$, dans la limite lorsque la maille tend vers zéro, une interface démarrant aussi proche de $A$ que désiré existe presque sûrement. Les interfaces, parcourues en s'éloignant de $A$, ont la plage ouverte à leur droite et la fermée à leur gauche. Ces dessins ne tracent l'interface que jusqu'au moment où elle atteint soit le côté $BC$ soit le côté $CD$. L'information dessinée est effectivement suffisante pour dire qu'il n'y aura pas de traversée horizontale dans le dessin de gauche, alors qu'il y en a une dans celui de droite. En effet, dans le dessin de gauche, les deux parties les plus à droite de l'interface (celle allant d'un point de $AD$ à un de $AB$ et celle de $AB$ à $BC$) délimitent une plage de sites fermés qui bloque toute traversée. Au contraire, dans le dessin de droite, la dernière partie de l'interface, celle allant de $AB$ à $CD$ a, à sa droite, une plage de sites ouverts qui assure une traversée horinzontale. Ainsi, suivre l'interface démarrant au coin $A$ permet de décider si la configuration sous-jacente possède une traversée horizontale : elle en aura une si et seulement si elle atteint le côté $CD$ avant le côté $BC$.
\begin{center}
\begin{figure}[h!]
\includegraphics[width = .45\linewidth]{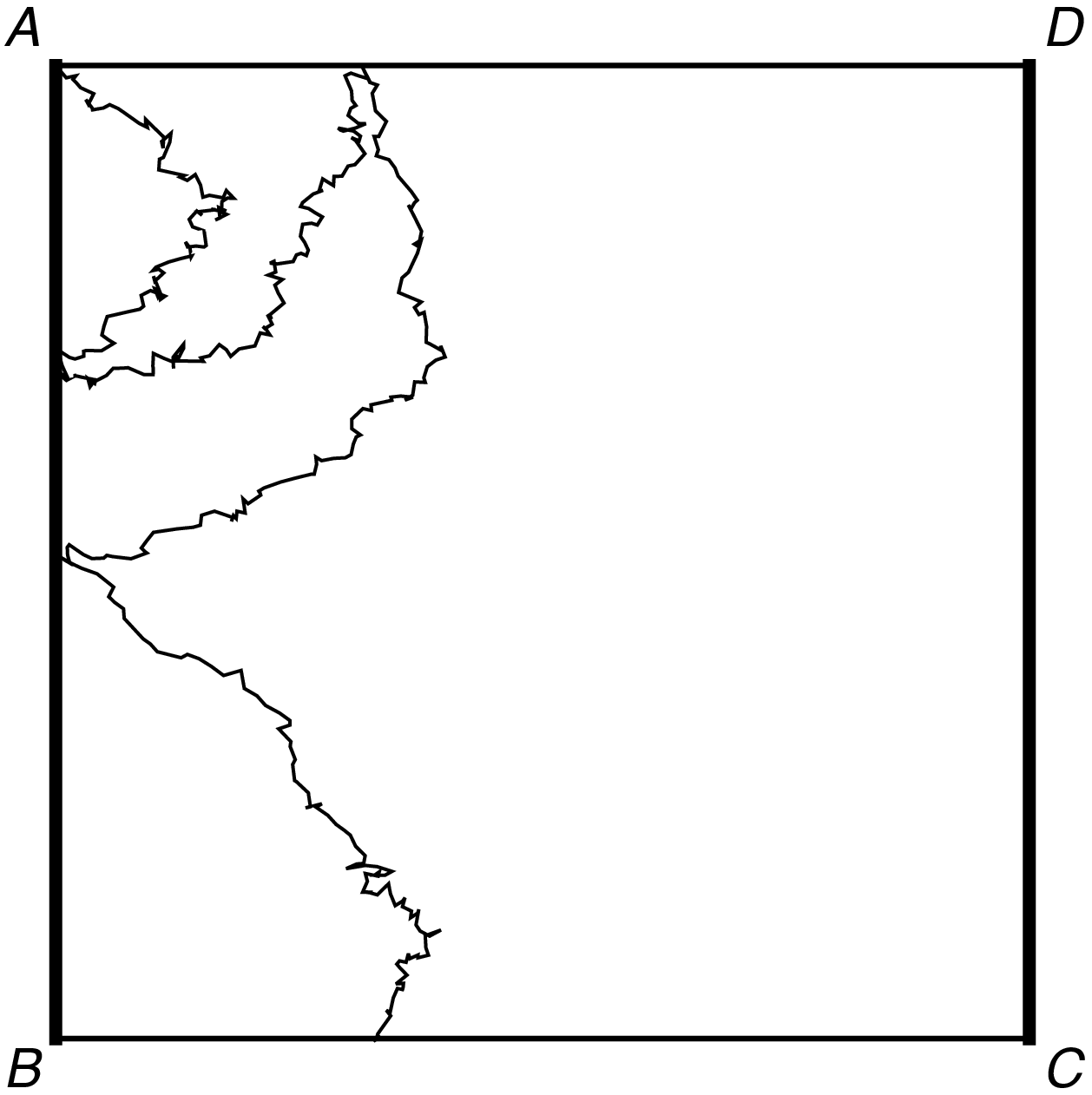}\hfill
\includegraphics[width = .45\linewidth]{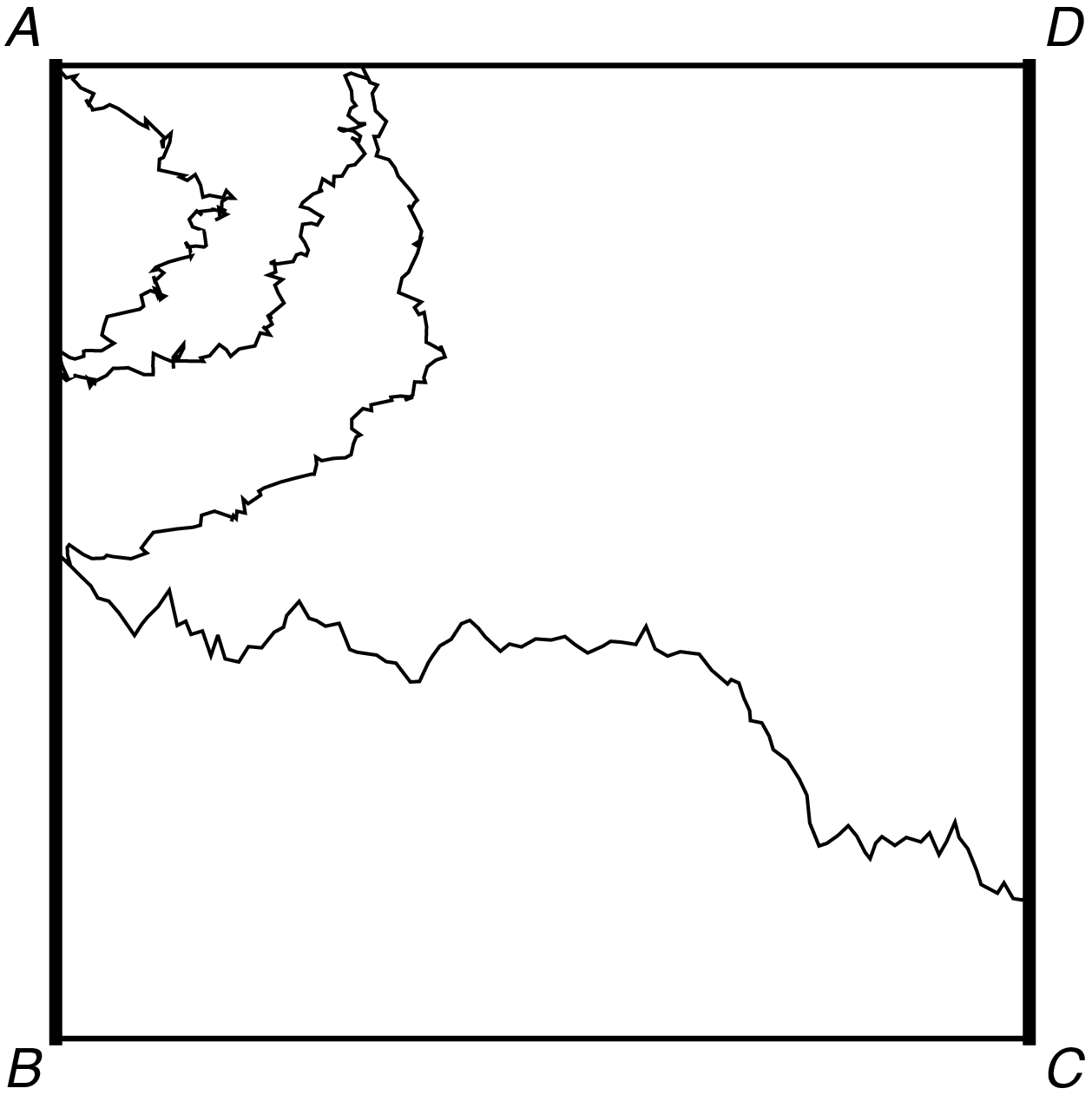}\\
\includegraphics[width = .45\linewidth]{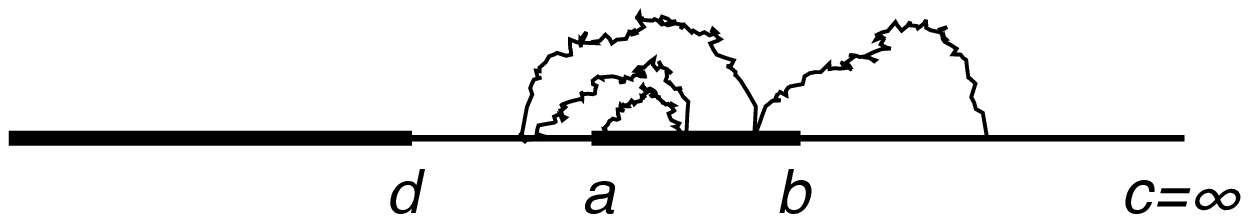}\hfill
\includegraphics[width = .45\linewidth]{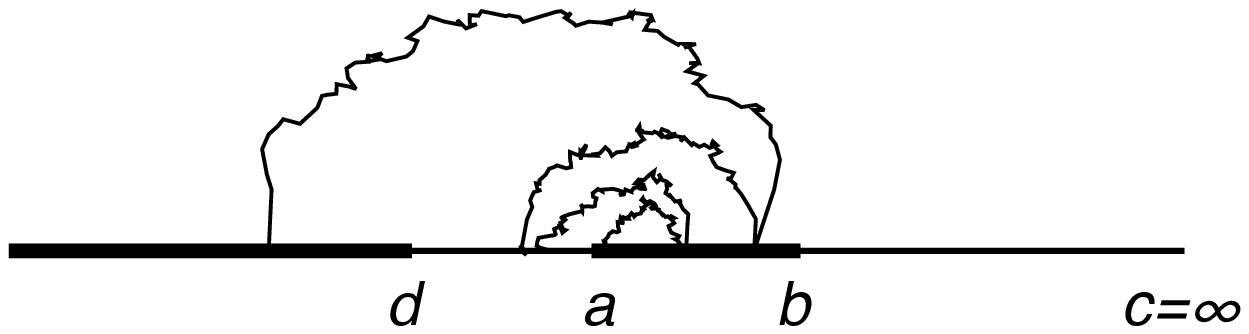}
\caption{Deux interfaces à l'intérieur d'un domaine carré (en haut) et leur image dans le demi-plan supérieur (en bas). Les lettres $a,b,c$ et $d$ étiquettent les images par $\phi$ des sommets $A,B,C$ et $D$.
}\label{fig:cinq}
\end{figure}
\end{center}
\noindent Soit $\phi$ l'inverse de l'application de Schwarz-Christoffel qui envoie l'intérieur d'un rectangle $ABCD$ sur le demi-plan supérieur. Les figures sous celles des domaines carrés (figure \ref{fig:cinq}) représentent l'image des côtés $AB$, $BC$, $CD$ et $DA$ le long de la frontière horizontale du demi-plan, ainsi que les méandres de l'interface tracée dans le carré au-dessus. Notons par la lettre minuscule correspondante l'image par $\phi$ des sommets $A$, $B$, $C$ et $D$. L'observation qu'une traversée dans le domaine carré existe si et seulement si l'interface atteint le côté $CD$ avant le côté $BC$ se traduit, pour les \og traversées \fg{} dans le demi-plan supérieur comme suit : une \og traversée \fg{} du segment $ab$ au segment allant de $-\infty$ à $d$ existera si et seulement si l'interface partant de $a$ atteint la demi-droite de $-\infty$ à $d$ avant qu'elle n'atteigne la demi-droite allant de $b$ à $+\infty$. Le problème de déterminer la probabilité de traversée horizontale (ou de tout événement décrit par le triplet $(D,I,J)$) est donc équivalent à déterminer la probabilité que l'interface partant de $a$ atteigne la demi-droite $-\infty$ à $d$ avant l'autre demi-droite.

\begin{center}
\begin{figure}[h!]
\includegraphics[width = .45\linewidth]{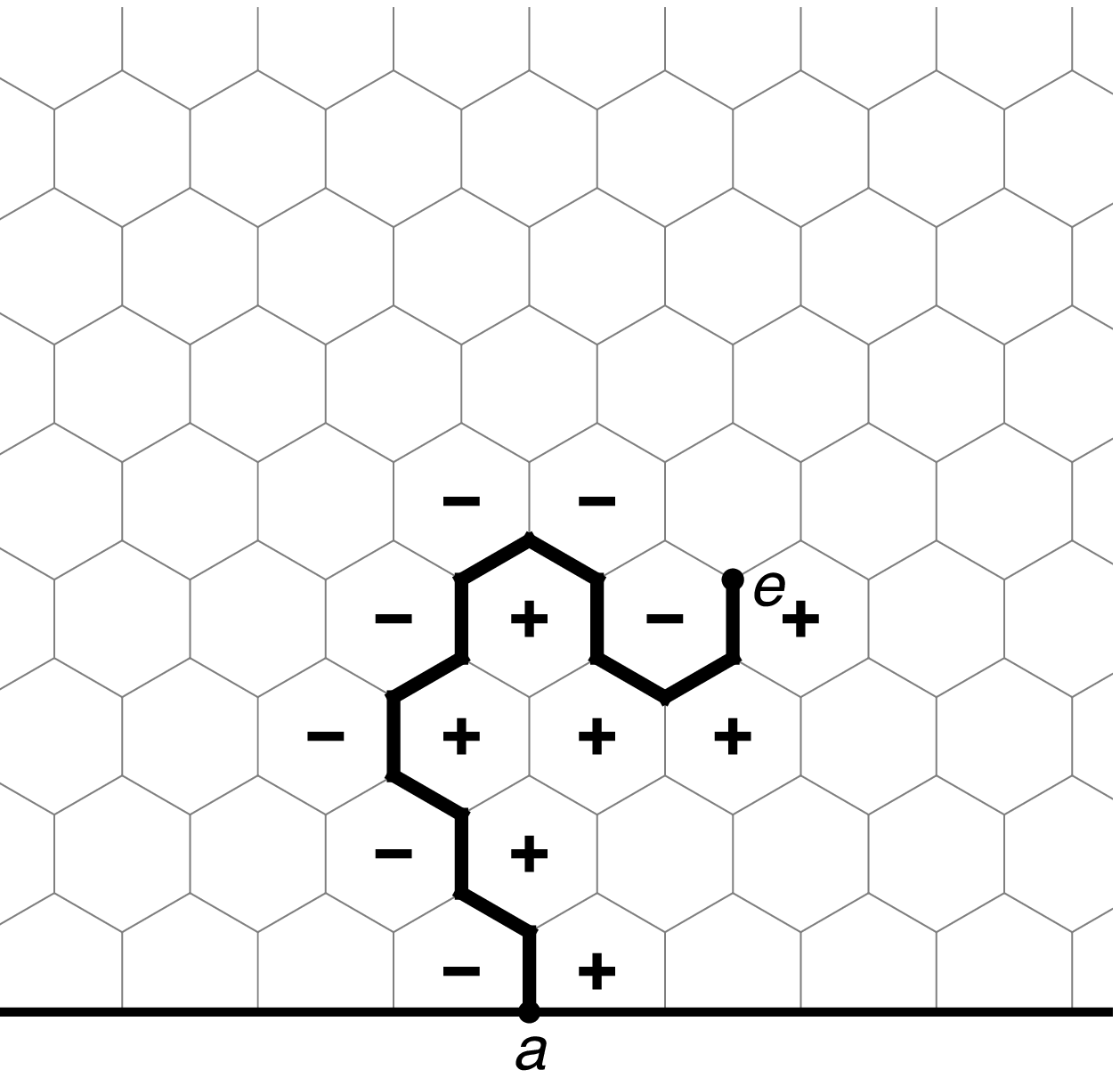}\hfil
\includegraphics[width = .45\linewidth]{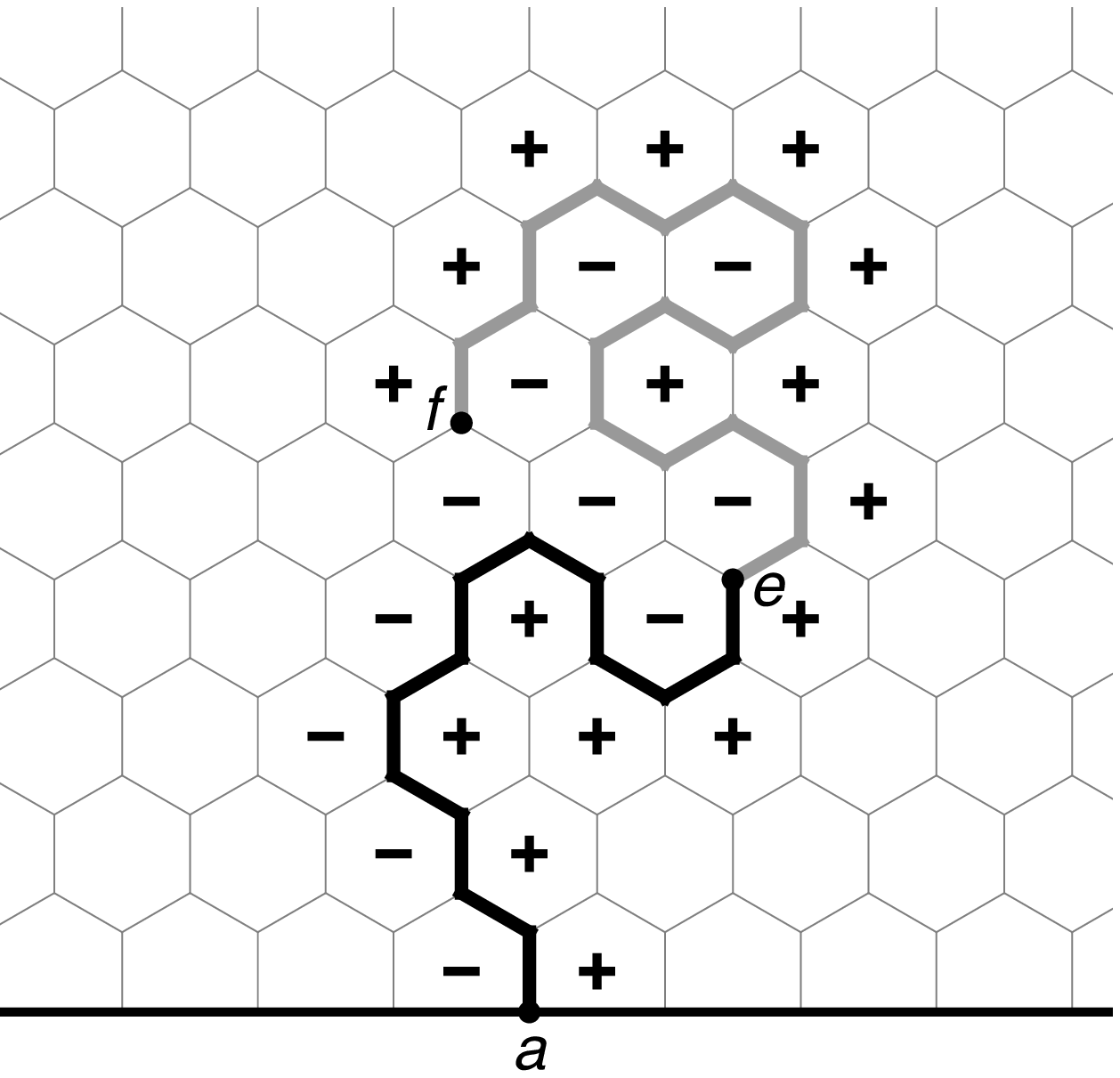}
\caption{La croissance d'une interface de l'origine $a$ à un point $e$ (à gauche), puis jusqu'à $f$ (à droite).}\label{fig:six}
\end{figure}
\end{center}
L'argument précédent permet de réduire l'étude des probabilités de traversée à celle des probabilités des interfaces attachées à l'origine du demi-plan complexe $\mathbb D$. Quoique basé sur l'hypothèse d'invariance conforme, il ne révèle pas comment cette hypothèse relie les probabilités de diverses interfaces. À cette fin, il est utile de considérer l'interface, non pas tracée dans sa totalité, mais plutôt comme un processus dynamique où elle croît à partir de cette origine, c'est-à-dire du point $a$. Cette interface croît en séparant les sites ouverts immédiatement à sa droite, marqués d'un $+$, et les fermés à sa gauche, marqués d'un $-$. La figure \ref{fig:six} représente deux captures de ce processus sur un réseau triangulaire où chaque site est représenté par un hexagone. Dans la figure de gauche, l'interface a cru jusqu'au point $e$. Pour décider de la prochaine étape de cette croissance, l'état, ouvert ou fermé, de l'hexagone du site \og devant \fg{} $e$ doit être choisi. Si cet hexagone est ouvert ($+$), l'interface ira vers la gauche et, s'il est fermé ($-$), vers la droite. Ce processus de croissance est facile à comprendre pour la percolation puisque l'état des sites est décidé indépendamment des autres sites. Cette indépendance est perdue dans d'autres systèmes physiques tel le modèle d'Ising qui sera discuté à la prochaine section; heureusement cette indépendance probabiliste n'est pas nécessaire aux arguments qui suivent ci-dessous. Le diagramme droit de la figure capture la croissance quelques instants plus tard. L'hexagone devant $e$ est fermé et l'interface a bifurqué vers la droite, puis s'est rabattue vers la gauche. Du point $f$, il ne sera pas nécessaire de choisir l'état de l'hexagone devant $f$ puisque ce site a déjà été fixé comme étant fermé à une étape précédente. Ainsi, même si le processus de croissance est simple, il dépend de toute l'histoire des choix précédents et est donc non-markovien. L'interface est appelée un {\em chemin auto-évitant}.

\begin{center}
\begin{figure}[h!]
\includegraphics[width = .30\linewidth]{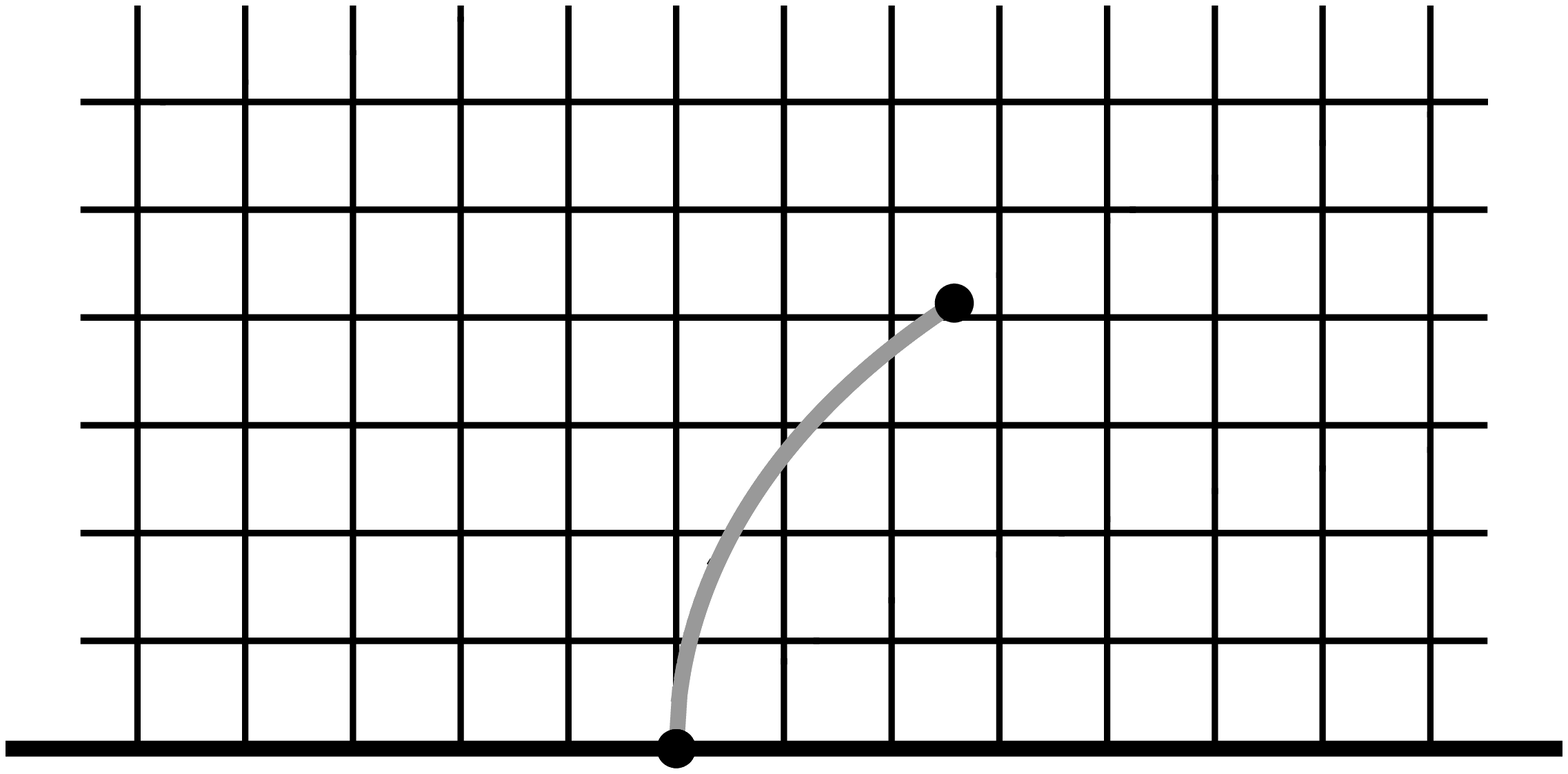}\hfill
\includegraphics[width = .30\linewidth]{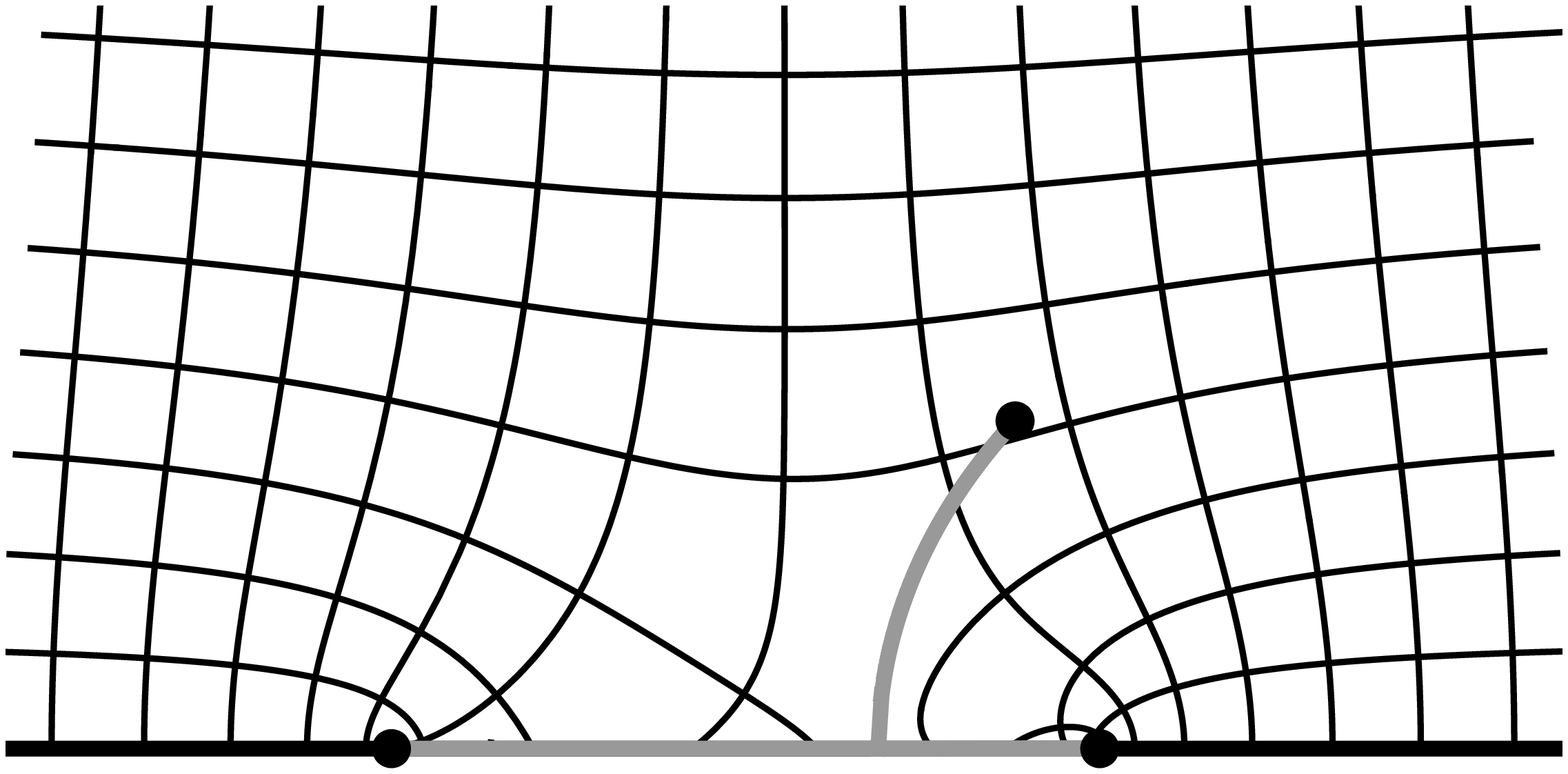}\hfill
\includegraphics[width = .30\linewidth]{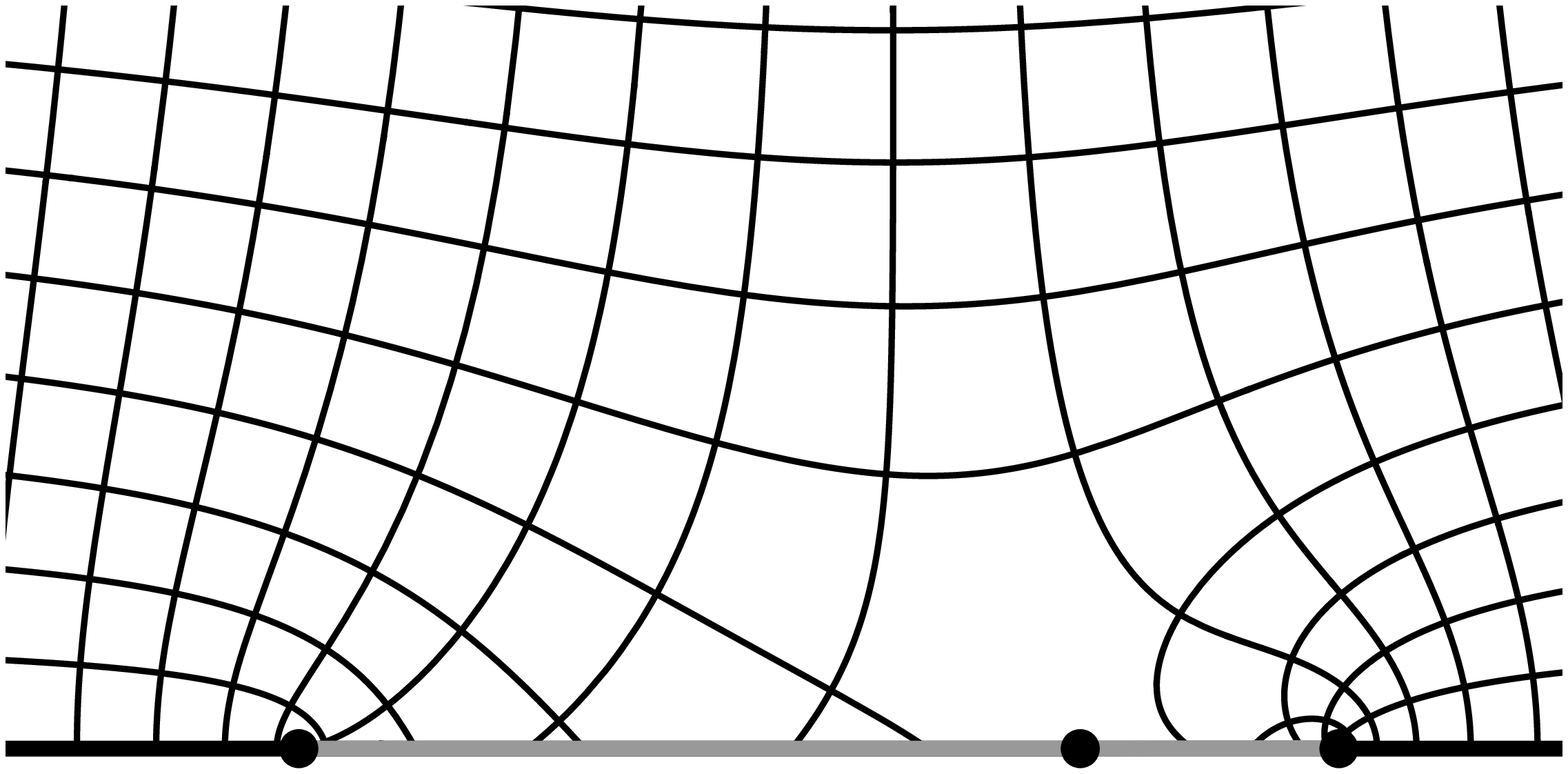}
\caption{Le demi-plan supérieur duquel une interface a été retirée est envoyé holomorphiquement sur le demi-plan supérieur.
}\label{fig:sept}
\end{figure}
\end{center}
Cette propriété non-markovienne a une conséquence simple : l'interface croît à partir de $f$ dans le demi-plan supérieur $\mathbb D$ duquel a été retirée l'interface de l'origine jusqu'à ce point $f$. La frontière de cette région va de $-\infty$ à l'origine $a$, suit alors le côté gauche de l'interface jusqu'à $f$, puis revient par son côté droit à l'axe horizontal qu'elle suit alors jusqu'à $+\infty$. Si l'interface est continue, cette région est un sous-ensemble ouvert simplement connexe du demi-plan et, par le théorème de Riemann, il existe donc une bijection holomorphe qui l'envoie sur ce même demi-plan. Si, de plus, un paramétrage $\gamma:[0,t_f]\to \mathbb D$ de l'interface est choisi, avec $\gamma(0)=a$ et $\gamma(t_f)=f$, alors il existe une famille de fonctions holomorphes $\psi_t, 0\leq t\leq t_f$, chacune \og retirant \fg{} la partie $\gamma([0,t])$ de l'interface de l'origine jusqu'à $\gamma(t)$. La figure \ref{fig:sept} capture trois étapes de ce processus pour une interface (ou fente) très simple : la première montre l'interface complète et la dernière l'image de $\psi_{t_f}$ où l'interface a été complètement absorbée par l'axe réel. Des droites avec partie réelle ou imaginaire constante (et leur image) ont été tracées. Ces courbes montrent que les angles sont conservés par les fonctions $\psi_{t}$. L'extrémité $f$ et l'origine de la fente (ici dédoublée puisqu'elle apparaît sur les deux côtés de cette interface) sont marquées par des points. Enfin l'interface elle-même, ou son image sous les $\psi_t$, est tracée en gris. Ces points et l'interface montrent la \og fonte \fg{} progressive de l'interface dans l'axe réel. Cette famille $\psi_t$ de fonctions holomorphes satisfait $\psi_{t=0}(z)=z$ pour tout $z\in\mathbb D$ et est uniquement déterminée si la condition suivante sur le comportement à l'infini est ajoutée : $\psi_t(z)=z+2t/z+\mathcal O(z^{-2})$. Cette dernière condition fixe un paramétrage $\gamma(t)$ privilégié sur l'interface qui sera maintenant utilisé.

Cette bijection entre interfaces dans le demi-plan complexe et familles $\psi_t$ de fonctions holomorphes peut être poussée un cran plus loin. L'{\em équation différentielle de Loewner} 
$$\frac{\partial\ }{\partial t}\psi_t=\frac{2}{\psi_t-\xi(t)}$$
lie cette famille $\psi_t$ à une fonction (continue) réelle $\xi:[0,t_f]\to\mathbb R$. Le point $\xi(t)$ est l'image du point $\gamma(t)$ de l'interface par $\psi_t$, c'est-à-dire $\psi_t(\gamma(t))=\xi(t)\in\mathbb R$. C'est donc le point de rencontre avec l'axe réel de la partie restante de l'interface au temps $t$. En inversant le paramètre d'évolution $t\to -t$, la famille $\psi_t$ peut être vue comme donnant naissance à une interface $\gamma$ à partir du point $\xi$. Ainsi cette fonction est parfois nommée la {\em force motrice} ({\em driving force}) puisque ses mouvements le long de l'axe horizontal détermineront l'apparence de $\gamma$. Évidemment, si seule l'interface $\gamma([0,t_f])$ est donnée, ni la famille $\psi_t$, c'est-à-dire l'évolution de Loewner, ni la fonction $\xi(t)$ ne peuvent être aisément déterminées à partir de l'équation ci-dessus. L'importance de l'équation de Loewner réside plutôt dans le fait qu'elle établit une bijection entre les interfaces dans le demi-plan supérieur et les fonctions continues $\xi$ décrivant le mouvement du point moteur sur l'axe réel auquel est attachée l'interface alors qu'elle est absorbée par cet axe (ou qu'elle en émane, si le paramètre $t$ est inversé).

Cette façon d'appréhender les interfaces, par le processus dynamique de leur croissance ou par l'évolution de Loewner qui les fait disparaître, permet une formulation alternative de l'hypothèse d'invariance conforme. Supposons à nouveau qu'un espace de probabilité $\Omega$ existe dont les éléments sont les limites, lorsque la maille $\delta$ du réseau tend vers $0$, des interfaces (infinies) construites comme à la figure \ref{fig:six}. Notons par $\gamma:[0,+\infty)\to \mathbb D$ ces interfaces. Soit un ensemble mesurable $\omega$ de telles interfaces partageant leur partie initiale, disons jusqu'en un point $e$ fixé atteint au temps $t_e$. Ainsi, si $\gamma_1$ et $\gamma_2$ appartiennent à $\omega$, alors $\gamma_1(t)=\gamma_2(t)$ pour tout $t\in[0,t_e]$. Soit enfin $\psi_{t_e}$ la fonction holomorphe retirant la partie commune $\gamma([0,t_e])$ de ces interfaces. Alors l'hypothèse d'invariance conforme requiert que la mesure de l'ensemble $\{\psi_{t_e}(\gamma)\,|\,\gamma\in\omega\}$ soit égale à celle de $\omega$.

L'hypothèse d'invariance conforme contraint donc les espaces de probabilité $\Omega$. En fait, la bijection obtenue par l'équation de Loewner entre les interfaces $\gamma:[0,+\infty)\to\mathbb D$ et les fonction $\xi:[0,+\infty)\to \mathbb R$ permet de les caractériser complètement. Si $\Omega$ est un tel espace de probabilité, alors il existe un nombre positif $\kappa>0$ tel que la mesure sur ces interfaces dans $\mathbb D$ est induite par la bijection de Loewner si celle sur les fonctions réelles $\xi(t)$ est donnée par $\sqrt\kappa B(t)$ où $B$ est le mouvement brownien.

C'est Schramm qui formula (rigoureusement) ces idées et démontra le théorème de classification de ces espaces de probabilité de courbes conformément invariants \cite{schramm1,schramm2}. Il nomma le processus de croissance des interfaces l'{\em équation de Loewner stochastique}, mais la communauté changea ce nom pour {\em équation de Loewner-Schramm} préservant ainsi l'acronyme anglais {\em SLE} qu'il avait choisi. En utilisant un résultat de Stanislav Smirnov \cite{Smirnov} prouvant que les interfaces de la percolation sur un réseau triangulaire donnent lieu, dans la limite $\delta\to 0$, à l'espace $\Omega$ correspondant à $\kappa=6$, Schramm put démontrer la formule de Cardy. L'équipe constituée de Greg Lawler, Oded Schramm et Wendelin Werner, à laquelle se joignirent leurs collègues et étudiants, développa ses idées en une théorie élégante, démontrant rigoureusement de nombreux résultats prédits par l'intuition des physiciens et, surtout, introduisant des outils mathématiques pour attaquer plusieurs autres questions physiques liées aux transitions de phase. Werner obtint la Médaille Fields en 2006 pour ses contributions à ce corpus.

\end{section}


\begin{section}{L'invariance conforme du modèle d'Ising}

Malgré la beauté des idées de Schramm et du grand nombre de résultats qui en découlèrent, une preuve de l'invariance conforme de la limite continue des modèles physiques existait que dans très peu de cas. À la publication du premier article de Schramm \cite{schramm1}, il n'y avait à ma connaissance que la preuve de Smirnov pour la percolation sur un réseau triangulaire. Cette preuve, écrite dans un style compact, ne semblait pas pouvoir être étendue aisément à d'autres modèles. Il est donc naturel de se demander quels autres modèles sur réseau possèdent une limite décrite par l'espace de probabilité $\Omega$, non seulement pour $\kappa=6$, mais aussi pour toute autre valeur permise de ce paramètre. Les hypothèses d'invariance conforme et d'universalité peuvent être étendues à ces autres valeurs. Le modèle d'Ising fournit un exemple d'une telle limite qui, elle, va vers l'espace $\Omega$ en ${\kappa=3}$. Le but de cette section, modeste comme celui de la précédente, est de présenter les outils intervenant dans la preuve de Smirnov \cite{Smirnov2} de l'invariance conforme de ce modèle statistique sur le réseau carré. 

Il existe deux présentations du modèle d'Ising. La première que je présente d'abord est la définition originale et certainement la plus connue. J'introduirai après la seconde que Smirnov utilise dans sa preuve.

Soit $R$ un rectangle contenant $M\times N$ points de $\mathbb Z^2$. À chacun de ces points est attachée une variable aléatoire $\sigma_p$, $p=(i,j)$, prenant les valeurs $+1$ ou $-1$. À un choix $\sigma=(\sigma_{p=(i,j)})_{1\leq i\leq M,1\leq j\leq N}$ de ces $MN$ variables, on associe une {\em énergie}
$$E(\sigma)=-J\sum_{\langle p,p'\rangle}\sigma_p\sigma_{p'}-H\sum_p \sigma_p$$
où la seconde somme porte sur tous les points $p$ du rectangle $R$ et la première sur les paires $\langle p,p'\rangle$ de voisins immédiats : si $p=(i,j)$, alors $p'\in \{(i+1,j),(i-1,j),(i,j+1),(i,j-1)\}\cap R$. Les constantes $J$ et $H$ ont une interprétation physique : $J$ décrit le couplage ferromagnétique entre les sites ($=$ spins) voisins et $H$ est lié au champ magnétique extérieur dans lequel baignent les spins. L'{\em ensemble des configurations} $\{\sigma\}=\mathbb Z_2^{M\times N}$ est pourvu d'une mesure $P(\sigma)=e^{-E(\sigma)/kT}/Z$ où $Z$ est un facteur de normalisation assurant que $\sum_{\sigma}P(\sigma)=1$ et $kT$ est le produit de la constante de Boltzmann et de la température $T$. On notera que seuls les rapports $J/kT$ et $H/kT$ interviennent; il est donc usuel de remplacer $J/kT$ par $\beta$ et $H/kT$ par une autre constante $\eta$. Le modèle d'Ising est donc la famille d'espaces de probabilité $(\mathbb Z_2^{M\times N}, P_{\beta,\eta})$ étiquetée par l'inverse de la température $\beta$ et la constante $\eta$. J'ai présenté le modèle sur un rectangle $\subset \mathbb Z^2$, mais la définition s'étend aisément à des (hyper-)rectangles dans $\mathbb Z^d$ pour $d\geq 1$. (La description du modèle en $d=3$ demeure à ce jour un domaine de recherche actif.) Des conditions aux limites peuvent être ajoutées, par exemple en identifiant les deux côtés verticaux du rectangle $R$. Enfin, la limite de la maille du réseau est ici entendue comme une limite où le nombre de sites $M\times N$ tend vers l'infini.

Notons que, si la constante $J$ est positive, alors une configuration $\sigma$ est d'autant plus probable que le nombre de paires $\langle p,p'\rangle$ où les spins coïncident est grand. Cet effet sera d'autant plus marqué que la température $T>0$ sera petite. Ainsi, lorsque $J$ est positive, les spins tendent à s'orienter dans la même direction.  Si $H$ est positif, les configurations $\sigma$ avec un grand nombre de sites $+1$ seront favorisées. Le champ extérieur $H$ force donc les spins à s'aligner dans la direction (positive ou négative) où il pointe. Pour la suite, le champ extérieur $H$ sera posé à zéro.

Il a été utile de concevoir les interfaces en percolation comme un processus de croissance dynamique où chaque site est décidé indépendamment des précédents. Cette façon d'appréhender les interfaces ne fonctionne pas dans le cas des interfaces entre les plages avec sites $+1$ et celles avec $-1$. En effet, ici, les variables aléatoires $\sigma_p$ et $\sigma_{p'}$ ne sont pas indépendantes sous la mesure $P(\sigma)$. Les arguments de Schramm menant aux espaces $\Omega$ ne requièrent cependant pas cette indépendance.

La physique du modèle statistique est révélée par l'espérance de certaines variables, par exemple celle de l'énergie par site $E(\sigma)/MN$ ou par celle des fonctions de corrélations à $n$ points, c'est-à-dire des produits $\sigma_{p_1}\sigma_{p_2}\dots\sigma_{p_n}$ pour des points $p_i\in R$ distincts. Le signe révélateur d'une transition de phase pour un modèle physique comme le modèle d'Ising est la singularité d'une de ces valeurs moyennes ou d'une de leurs dérivées. Le calcul exact de l'énergie par site en $d=2$ par Lars Onsager en 1944 démontra l'existence d'une telle singularité en un certain $\beta_c$, $0<\beta_c<\infty$ en $H=0$. (La valeur de l'inverse de la température critique est donnée implicitement par $\sinh 2\beta_c=1$.) Ce calcul, considéré comme un tour de force, prouve donc l'existence d'une transition de phase pour le modèle d'Ising en deux dimensions.

Plutôt que d'utiliser un sous-ensemble rectangulaire de $\mathbb Z^2$ dont la base est horizontale, la seconde présentation opte pour un rectangle incliné à 45$^\circ$. Le diagramme de gauche de la figure \ref{fig:huit} montre une configuration $\sigma$ où les sites $p$ sont occupés par le signe $\sigma_p$ et les paires de voisins immédiats sont liés par une diagonale en tirets. Une {\em configuration de boucles} sera maintenant construite. La première étape consiste à remplacer toutes les diagonales liant deux signes opposés par une tuile contenant deux quarts de cercle positionnés pour que ces arcs coupent la diagonale qu'ils remplacent. Le second diagramme de la figure \ref{fig:huit} montre le résultat. La seconde étape remplace maintenant les diagonales liant des signes identiques. Les quarts de cercle pourront être positionnés pour couper la diagonale ou non. Ils la couperont avec probabilité $q=e^{-2\beta}$ et ne la couperont pas avec probabilité $p=1-q$. En ajoutant les demi-cercles à la frontière comme indiqué sur le dernier diagramme de la figure \ref{fig:huit}, la configuration ainsi obtenue ne contient que des boucles fermées. Fortuin et Kasteleyn \cite{fortuin} ont montré qu'une telle configuration de boucles $\gamma$ apparaît alors avec la probabilité $q^{c(\gamma)}p^{d(\gamma)}2^{b(\gamma)}/Z$ où $c(\gamma)$ et $d(\gamma)$ sont respectivement les nombres de tuiles dont la diagonale est coupée ou préservée par les quarts de cercle, et où $b(\gamma)$ est le nombre de boucles fermées. (Pour la configuration de la figure \ref{fig:huit}, $b=9$, $c=26$ et $d=22$, le nombre total de tuiles étant $c+d=48$.) Le facteur de normalisation $Z$ est alors identique à celui de la première présentation. Les valeurs critiques de $p$ et $q$, c'est-à-dire celles en $\beta_c$, sont $2\sqrt2$ et $1-\sqrt2$. En ces valeurs, une simplification remarquable a lieu pour les domaines tel celui considéré : la probabilité d'une configuration de boucles $\gamma$ devient proportionnelle à $(\sqrt2)^{b(\gamma)}$. C'est donc cet espace de probabilité qu'utilise Smirnov dans sa preuve : si $R$ est un domaine connexe contenant $N_R$ tuiles, l'espace des configurations est $\mathbb Z_2^{N_R}$ et la mesure d'une configuration, c'est-à-dire d'un choix parmi $\{\,
\begin{tikzpicture}[baseline={1.0},scale=0.25]
\draw[line width=0.2mm] (0,0) -- (1,0) -- (1,1) -- (0,1) -- (0,0);
\draw[line width=0.2mm]  (0.5,0) arc (0:90:0.5);
\draw[line width=0.2mm]  (0.5,01) arc (180:270:0.5);
\end{tikzpicture},\
\begin{tikzpicture}[baseline={1.0},scale=0.25]
\draw[line width=0.2mm] (0,0) -- (1,0) -- (1,1) -- (0,1) -- (0,0);
\draw[line width=0.2mm]  (0.5,0) arc (180:90:0.5);
\draw[line width=0.2mm]  (0.5,01) arc (360:270:0.5);
\end{tikzpicture}\,
\}$ pour chaque tuile, est proportionnelle à $(\sqrt2)^{b(\gamma)}$.
\begin{center}
\begin{figure}[h!]
\includegraphics[width = .30\linewidth]{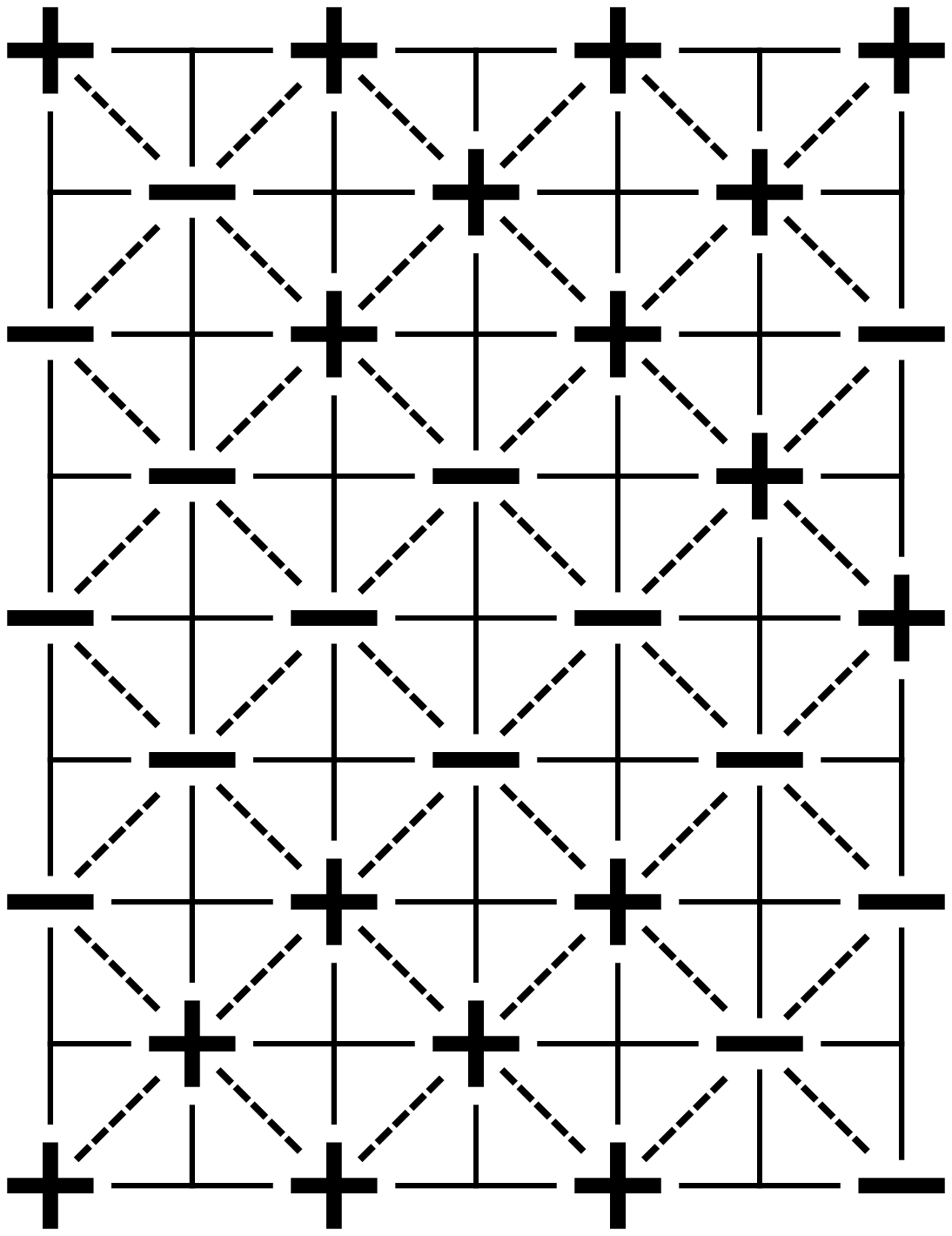}\hfill
\includegraphics[width = .30\linewidth]{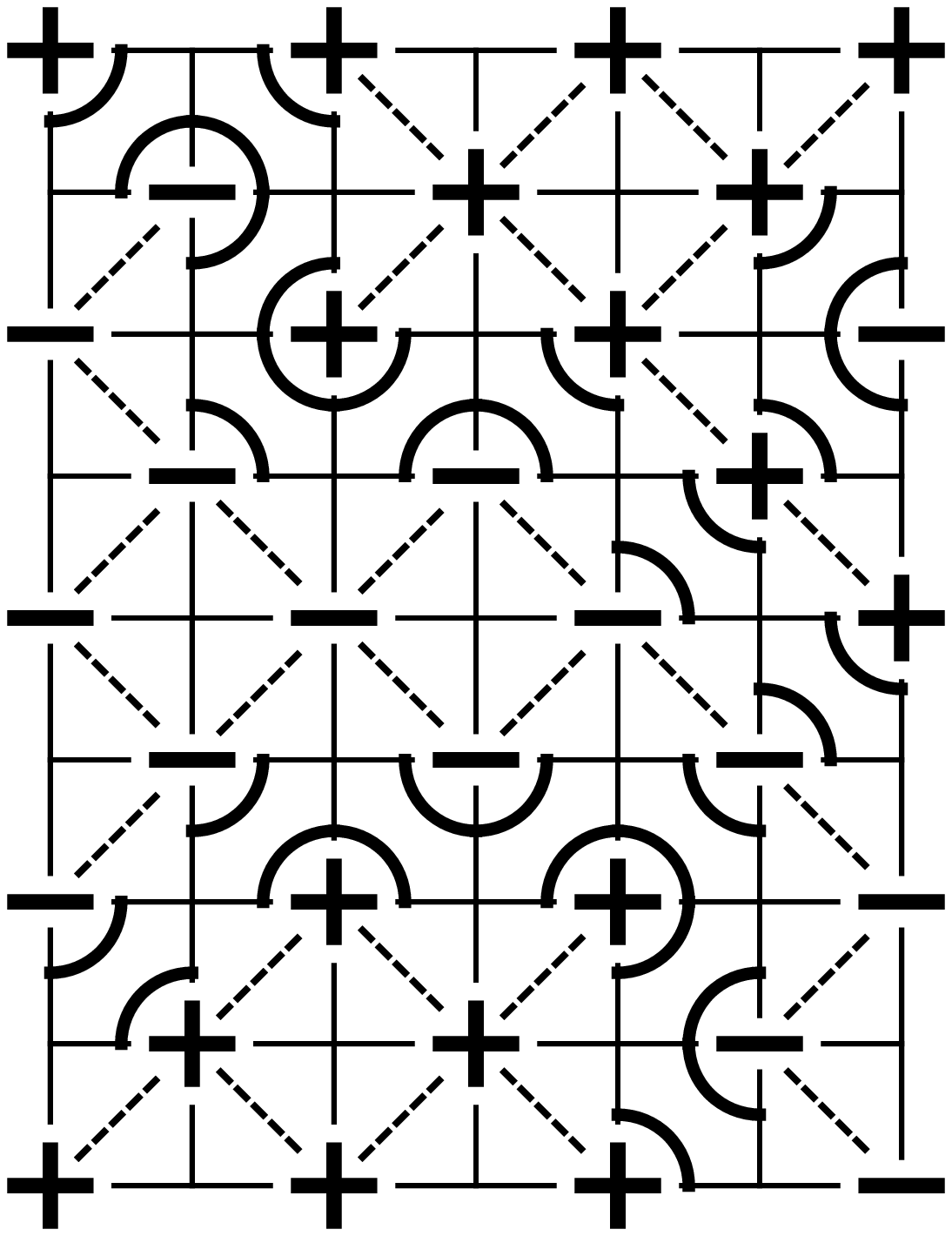}\hfill
\includegraphics[width = .30\linewidth]{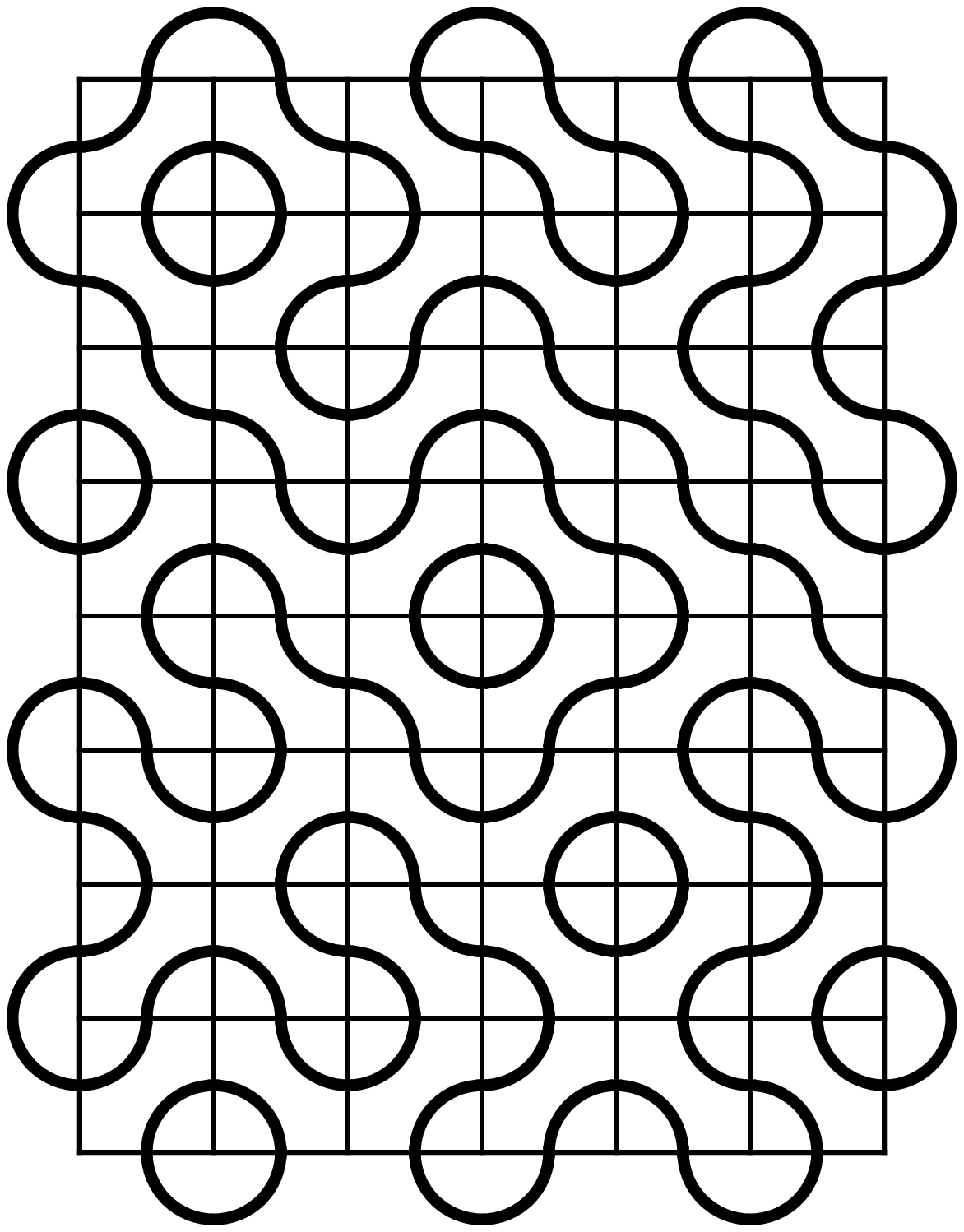}
\caption{D'une configuration de spins à une configuration de boucles.}\label{fig:huit}
\end{figure}
\end{center}

Il est ainsi possible de considérer un domaine quelconque du plan, comme celui représenté au diagramme gauche de la figure \ref{fig:neuf}. Les tuiles de la région $R$ sont maintenant celles qui intersectent l'intérieur de la courbe fermée. Des arcs frontières, tracés en gris, ont été ajoutés aux tuiles extérieures attenant au domaine $R$. Ces arcs ont été choisis de façon à ce que toute configuration $\gamma$ sur $R$ ne soit constituée que de boucles fermées (comme le diagramme droit de la figure l'indique) à l'exception d'une courbe entrant au point $a$ et sortant au point $b$ de $R$ qui est mise en évidence sur le diagramme droit.
\begin{center}
\begin{figure}[h!]
\includegraphics[width = .45\linewidth]{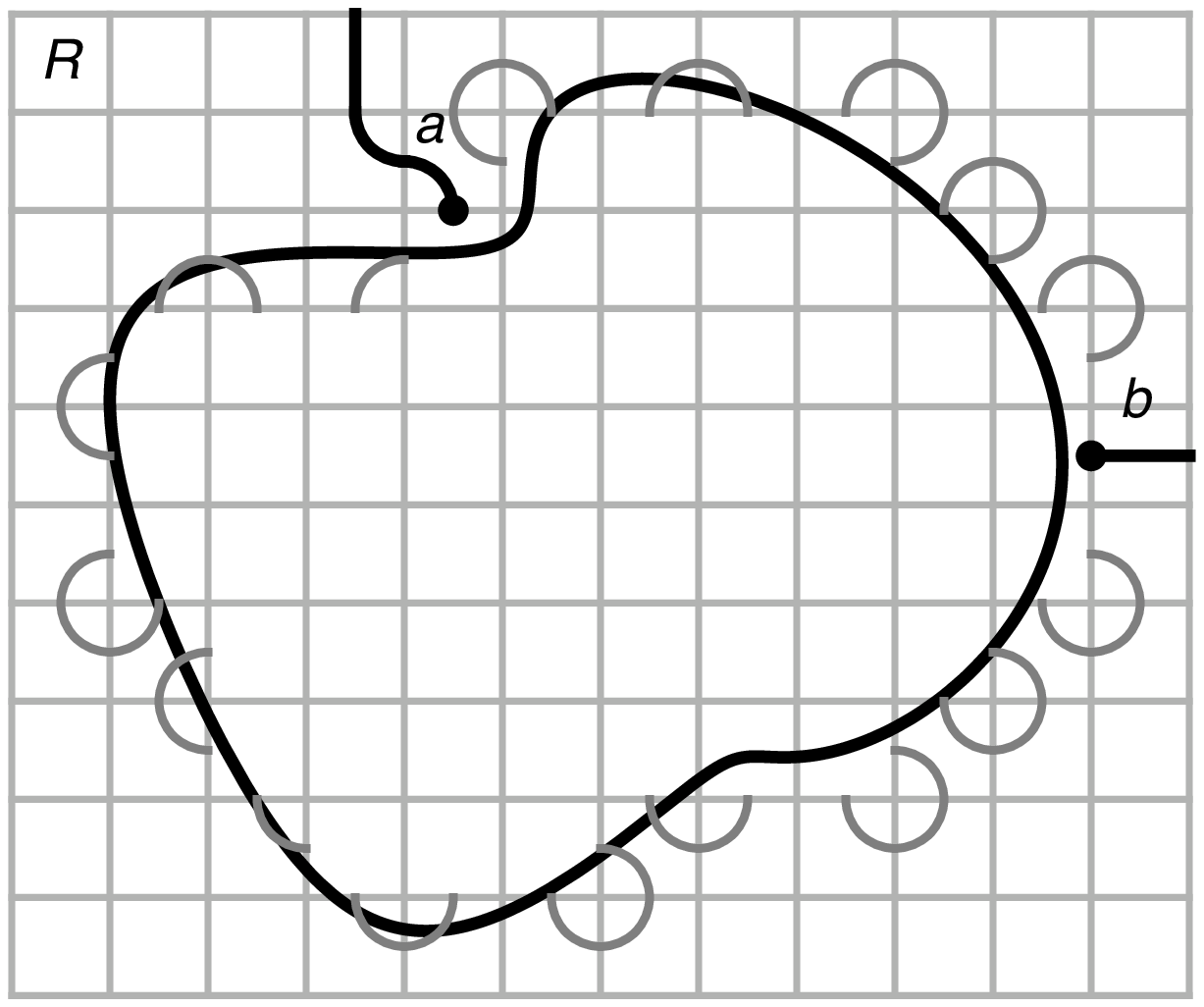}\hfill
\includegraphics[width = .45\linewidth]{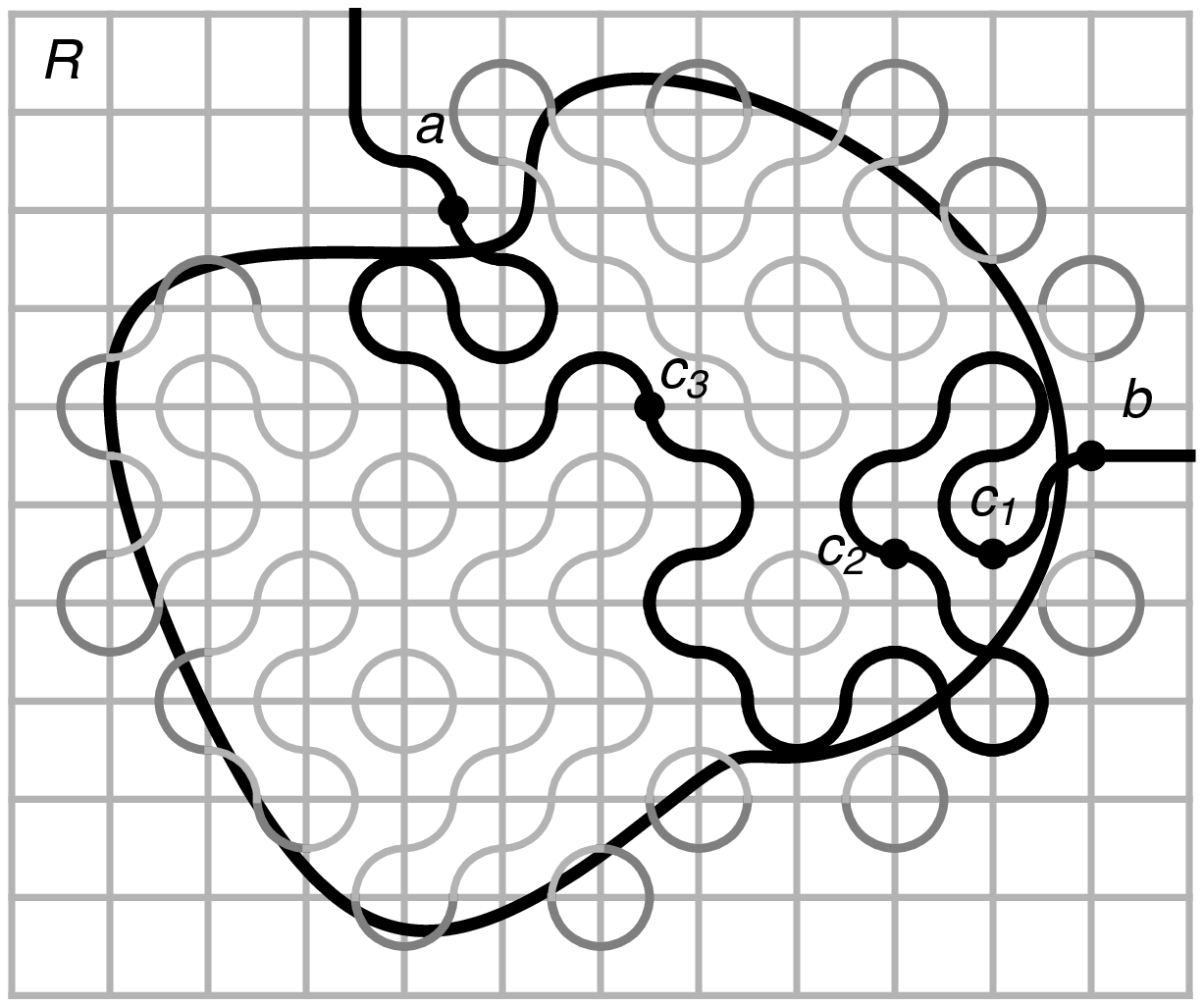}
\caption{\`A gauche un domaine $R$ et ses arcs frontières en gris; à droite une configuration de boucles, dans ce domaine et avec ces conditions aux limites, possédant une trajectoire allant des points frontières $a$ et $b$ avec trois points distingués sur cette trajectoire.}\label{fig:neuf}
\end{figure}
\end{center}

Le coeur de la preuve de Smirnov repose sur le choix d'une observable \og physique \fg{} pour laquelle il est possible de montrer que la limite existe et qu'elle est conformément invariante. Soit $C$ l'ensemble des côtés des tuiles qui sont à l'intérieur du domaine $R$.  Voici un premier choix pour cette observable, naturel quoique naïf. Soit $f:C\to [0,1]$ la fonction définie par l'espérance $f(c)$ que la trajectoire $\textrm{\it tr}(\gamma)$ de $a$ à $b$ passe par le côté $c\in C$. Clairement $f$ est une fonction réelle et, si elle possède une limite lorsque la maille du réseau tend vers zéro, cette limite sera également une fonction réelle. Si elle est holomorphe, elle sera une constante, ce qui n'est pas d'un grand intérêt physique. Ce premier choix doit être rejeté.

\begin{center}
\begin{figure}[h!]
\includegraphics[width = .30\linewidth]{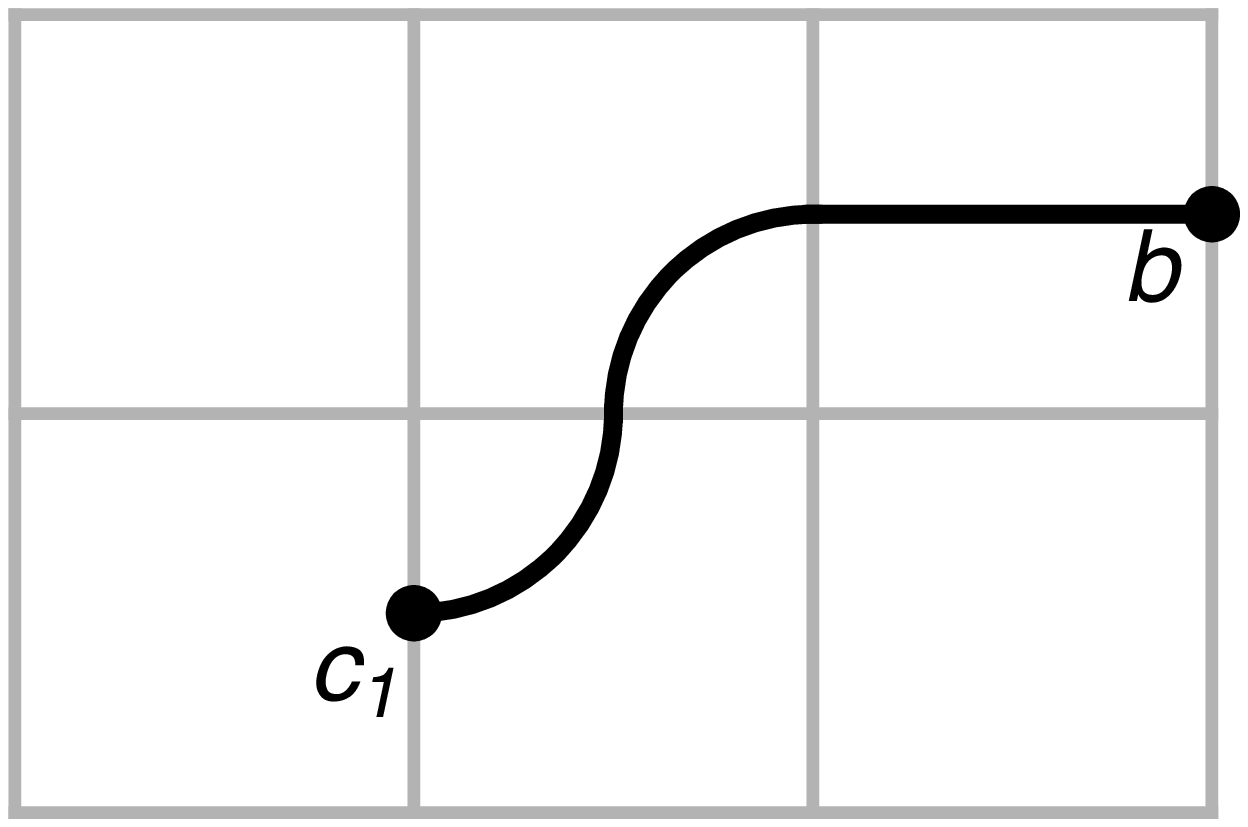}\hfill
\includegraphics[width = .30\linewidth]{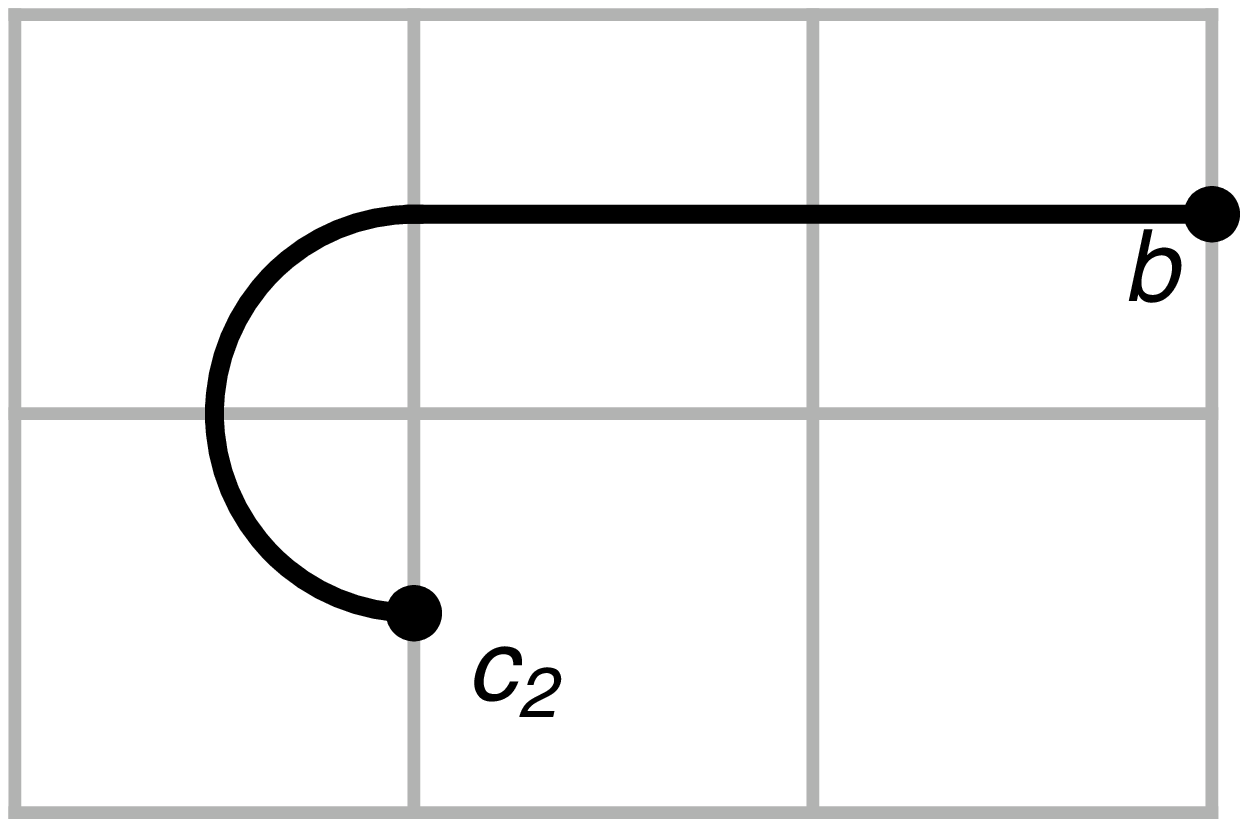}\hfill
\includegraphics[width = .30\linewidth]{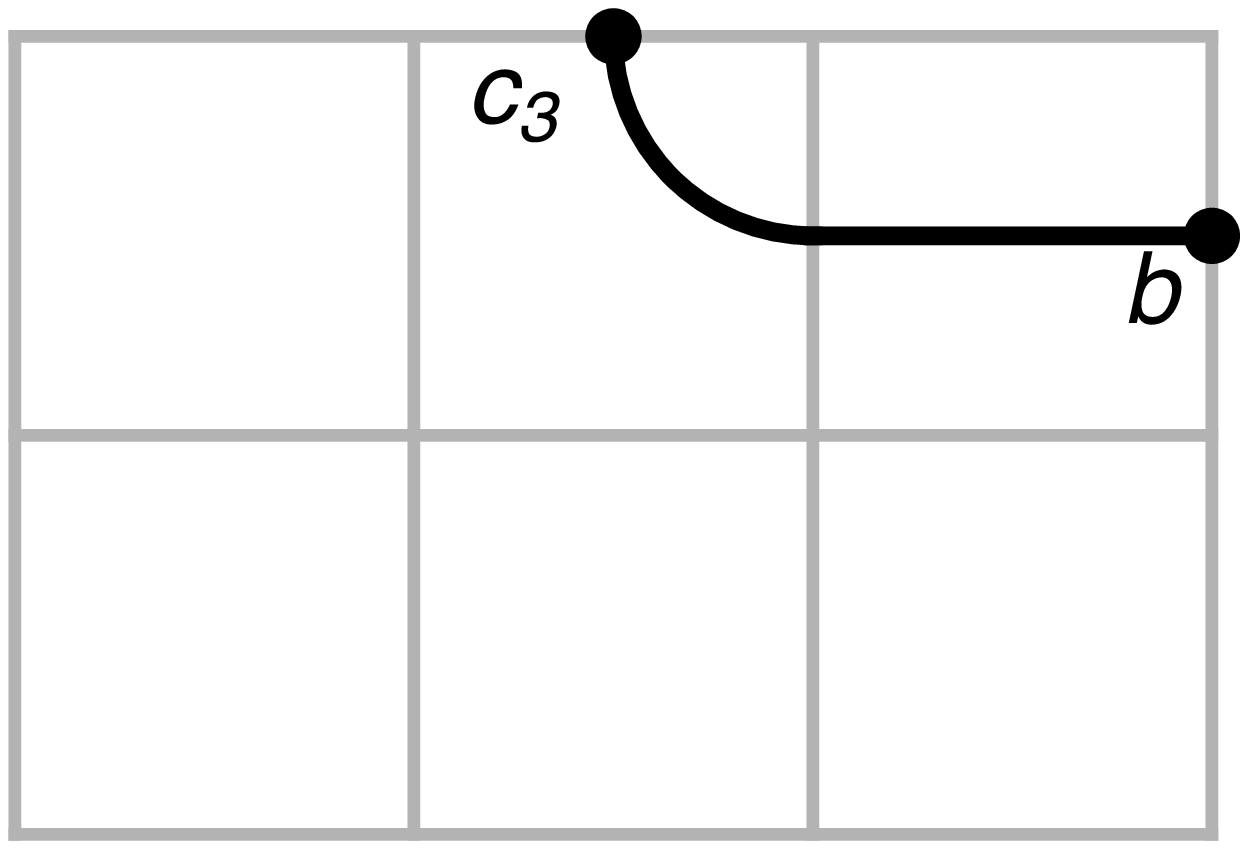}
\caption{Les accroissement $w(b\to c_i)$ de l'angle du vecteur tangent de $b$ aux points $c_1 ,c_2$ et $c_3$ sont $0, \pi$ et $-\pi/2$.}\label{fig:dix}
\end{figure}
\end{center}
L'idée de Smirnov est de pondérer cette fonction $f(c)$ en fonction de l'accroissement $w(b\to c)$ de l'angle que fait le vecteur tangent à la trajectoire en allant du point $b$ au point $c\in C$. Plus précisément soit $F:C\to \mathbb C$ la fonction donnée par $\mathbb E(\chi_{\textrm{\it tr}(\gamma)}(c)\cdot e^{-\frac{i\pi}2 w(b\to c)})$ où $\chi_{\textrm{\it tr}(\gamma)}$ est la fonction caractéristique sur la trajectoire allant de $a$ à $b$. Les accroissements $w(b\to c_i)$ de l'angle du vecteur tangent à la trajectoire jusqu'aux points $c_i$ peuvent être lus aisément de la figure \ref{fig:dix}. Ce poids, ajouté au parcours de la trajectoire, n'était pas inconnu des physiciens. Il est la clé du calcul purement combinatoire de la fonction de partition du modèle d'Ising fait par Kac et Ward \cite{kac}. (Voir également le chapitre 5 du cours de mécanique statistique de Feynman \cite{feynman}.) Il n'en demeure pas moins inspiré. Cette fonction $F$ de domaine $C$ prend maintenant ses valeurs dans $\mathbb C$, et non pas seulement dans $\mathbb R$, et son caractère holomorphe (discret) peut être étudié. Les équations de Cauchy-Riemann discrète se lisent
$$F(c_{no})-F(c_{se}) = i(F(c_{ne})-F(c_{so}))$$
où les arêtes $c_{no},c_{se},c_{ne}$ et $c_{so}$ sont celles contiguës à $c$ dans les directions nord-ouest, sud-est, et ainsi de suite. (Voir la figure \ref{fig:onze}.) Smirnov utilise une définition légèrement différente de cette discrétisation, mais il devra vérifier cette équation dans une preuve délicate et minutieuse. 
\begin{center}
\begin{figure}[h!]
\includegraphics[width = .350\linewidth]{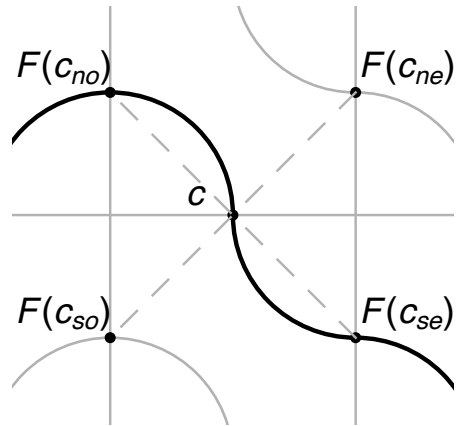}
\caption{Les arêtes contiguës à $c$ apparaissant dans l'équation de Cauchy-Riemann discrète.}\label{fig:onze}
\end{figure}
\end{center}
La définition de $F$ force également sa phase à la frontière du domaine $R$. En effet, si $z$ est un point de la frontière de $R$ ou très proche, l'espérance de l'accroissement $w(b\to z)$ sera, pour une maille du réseau très fine, mesurée avec un vecteur tangent à $\textrm{\it tr}(\gamma)$ arrivant en $z$ perpendiculairement à la frontière $\partial R$. Cette condition aux limites, avec l'équation de Cauchy-Riemann discrète détermine donc un problème aux limites de Riemann. La solution du problème continu correspondant est donnée par la fonction $(\Phi')^{\frac12}$ où $\Phi$ est l'application conforme envoyant le domaine $R$ sur le ruban horizontal infini de largeur unité, les points $a$ et $b$ étant envoyés aux extrémités.

Cette description ne fait que reformuler le problème de prouver l'invariance conforme du modèle d'Ising sur un réseau carré. De cette mise en place, la preuve de Smirnov procède alors en deux étapes : établir d'abord l'holomorphicité discrète de $F$, puis assurer que le passage à la limite peut être fait. Ces preuves établissent donc que, dans cette la limite, la fonction discrète $F$ définie sur les arêtes du réseau tend vers la fonction analytique $(\Phi')^{\frac12}$. Smirnov obtint la Médaille Fields en 2010 pour la preuve de l'invariance conforme pour les modèles de percolation et d'Ising en deux dimensions.

\end{section}


\begin{section}{Conclusion}

Les idées de Robert Langlands n'ont peut-être pas eu un impact aussi grand en physique mathématique qu'en théorie des formes automorphes. Il n'en demeure pas moins que, de par sa tribune privilégiée du {\it Bulletin of the American Mathematical Society}, son article {\it Conformal invariance in two-dimensional percolation} a permis de réorienter en partie les efforts de jeunes mathématiciens audacieux. Leurs travaux, et par conséquent les idées de Langlands, ont eu un impact important sur le domaine et de nombreux résultats ont suivi, réalisés par les équipes autour de Schramm et de Smirnov, mais aussi par la communauté mathématique. 

La classification de Schramm des espaces de probabilité satisfaisant l'hypothèse d'invariance conforme donne une réponse claire aux questions soulevées par notre hypothèse d'invariance conforme. Mais l'hypothèse d'universalité demeure une conjecture, même si l'invariance conforme de la limite de plusieurs autres modèles physiques a maintenant été démontrée. Il est à espérer que les années prochaines sauront énoncer précisément les hypothèses sous lesquelles cette seconde conjecture est vraie.

\end{section}


\section*{Remerciements}
L'auteur détient une subvention à la découverte du Conseil de recherches en sciences naturelles et en génie du Canada pour laquelle il est reconnaissant. 

\raggedright
\singlespacing


\end{document}